\documentclass{aa}  
\usepackage{graphicx}
\usepackage{txfonts}
\usepackage{lipsum}
\usepackage{subcaption}
\usepackage{lscape}
\usepackage{placeins}
\usepackage{hyperref}
\hypersetup{
  colorlinks=true,
  urlcolor=blue,
  linkcolor=blue,
  citecolor=blue
}
\usepackage{multirow}
\usepackage{xcolor}
\usepackage{soul}
\usepackage{CJKutf8}
\begin{document}
\begin{CJK*}{UTF8}{gbsn}
   \title{Consistency and precision of very long baseline interferometry source coordinate time series}
   \authorrunning{Zhang, X.-X., et al.}
   \titlerunning{A\&A proofs, manuscript no. aa61431-26}
   \author{X.-X. Zhang (张夏轩)\inst{1,2}
   \and N. Liu (刘牛)\inst{1,2}\fnmsep\thanks{\raggedright
  Corresponding authors:
  \texttt{niu.liu@nju.edu.cn};
  \texttt{sebastien.lambert@obspm.fr};
  \texttt{karine.lebail@chalmers.se}}
   \and S. B. Lambert\inst{3}\fnmsep\footnotemark[1]
   \and K. Le Bail\inst{4}\fnmsep\footnotemark[1]
   \and H. Kr\'asn\'a\inst{5}
   \and O. Titov\inst{6,7}
   \and M. Karbon\inst{8}
   \and T. Nilsson\inst{9,4}
   \and Z. Zhu (朱紫)\inst{1,2,10}
   \and J.-C. Liu (刘佳成)\inst{1,2}
   \and I. N. Huda\inst{11}
   \and J. Yang (杨军)\inst{4}
   \and X.-P. Cheng (程晓朋)\inst{12}
   \and B.-W. Sohn\inst{12,13}
   \and F.-C. Shu (舒逢春)\inst{6,14}
   \and X. He (何旋)\inst{6,15}
   \and Z.-W. Wang (王圳伟)\inst{1,2}
   \and Z.-Y. Zhang (张志云)\inst{1,2}
   \and H.-F. Yu (余华峰)\inst{1,2}
   \and J. Yao (姚俊)\inst{1,2}
   \and D.-D. Zhang (张大鼎)\inst{1,2}
   }

   \institute{School of Astronomy and Space Science, Nanjing University, Nanjing 210023, China
   \and Key Laboratory of Modern Astronomy and Astrophysics (Ministry of Education), Nanjing University, Nanjing 210023, China
   \and LTE, Observatoire de Paris, Universit\'e PSL, CNRS UMR8255, Sorbonne Universit\'e, Universit\'e de Lille, LNE, 61 avenue de l'Observatoire, 75014 Paris, France
   \and Department of Physics and Astronomy, Onsala Space Observatory, Chalmers University of Technology, 439 92 Onsala, Sweden
   \and TU Wien, Department of Geodesy and Geoinformation, Wiedner Hauptstra\ss e 8-10, 1040 Vienna, Austria
   \and Shanghai Astronomical Observatory, Chinese Academy of Sciences, Shanghai 200030, China
   \and Phase\&Rate, Canberra, Australia
   \and UAVAC, Dept. of Applied Mathematics and Aerospace Engineering, University of Alicante, Spain
   \and Lantm\"ateriet - The Swedish mapping, cadastral and land registration authority, Lantm\"aterigatan 2 C, 802 64 G\"avle, Sweden
   \and University of Chinese Academy of Sciences, Nanjing 211135, China
   \and Research Center for Computing, National Research and Innovation Agency, Bandung 40135, Indonesia
   \and Korea Astronomy and Space Science Institute, 776 Daedeok-daero, Yuseong-gu, Daejeon 34055, Republic of Korea
   \and University of Science and Technology, Gajeong-ro 217, Yuseong-gu, Daejeon 34113, Republic of Korea
   \and Shanghai Key Laboratory of Space Navigation and Positioning Techniques, Shanghai 200030, China
   \and University of Chinese Academy of Sciences, Beijing 100049, China
   }

   \date{Received 13 June 2026 / Accepted 15 September 2026}
  \abstract  
   {Coordinate time series of extragalactic radio sources produced by different analysis centers from very long baseline interferometry (VLBI) observations are widely used to evaluate source positional stability, select stable sources for celestial reference frames, assess the stability of reference-frame axes, and investigate source-related astrometric variability.
   The external consistency and realistic errors of these time-series products, as well as their relation to processing strategies and analysis configurations, remain to be further investigated.}
   {We aim to compare source coordinate time-series solutions from different analysis centers and investigate how processing strategies and configurations contribute to their differences.}
   {We collected eight coordinate time-series solutions from seven analysis centers and constructed a final sample of 496 common sources after data selection, reference-frame alignment, and the extraction of common observing sessions.
   The inter-solution differences were characterized using pairwise positional offsets and correlation analysis.
   The positional realistic errors of each solution were estimated with the N-cornered-hat (NCH) method and bootstrap resampling.}
   {For most solutions, the median weighted root mean square (WRMS) values of the coordinate time-series solutions are at the level of a few hundred micro-arc seconds $(\mathrm{\mu as})$, and the median NCH-derived precision is approximately $200$--$300\,\mathrm{\mu as}$ in right ascension and $250$--$400\,\mathrm{\mu as}$ in declination.
   Solutions with similar processing strategies, software packages, or source constraints typically show higher inter-solution consistency, as reflected by more similar WRMS values, stronger correlations, and closer NCH-derived precision estimates.
   The consistency among solutions is therefore affected not only by the broad distinction between global and independent modes but also by the detailed analysis configuration.
   The NCH-derived precision shows a clear dependence on source declination, with poorer precision in the southern sky, and becomes more stable for sources with a larger number of observing sessions.}
   {Different processing strategies and analysis configurations introduce marginal differences, while the source coordinate time-series solutions generally remain consistent.
   Inter-solution comparisons provide an independent way to obtain a realistic estimate of the stochastic errors of these products beyond their formal errors.
   }


   \maketitle
\section{Introduction}
\label{sec: Introduction}

The celestial reference frame (CRF) serves as the cornerstone of modern astrometry.
A precise CRF is essential for measuring positions of celestial bodies, determining Earth orientation parameters (EOPs), and tracking spacecraft in space.
At the current level of observational accuracy, distant extragalactic radio sources are generally assumed to be positionally fixed and are therefore suitable objects for realizing a quasi-inertial reference frame.
The International Celestial Reference System \citep[ICRS;][]{1998A&A...331L..33F}, adopted by the International Astronomical Union (IAU), provides the fundamental definition of the celestial reference system.
It is a quasi-inertial system centered at the Solar System's barycenter and nonrotating with respect to distant extragalactic objects.
The ICRS currently operates at radio wavelengths by the third realization of the International Celestial Reference Frame \citep[ICRF3;][]{2020A&A...644A.159C}, constructed from very long baseline interferometry (VLBI) observations of extragalactic radio sources \citep{1998AJ....116..516M}.
While ICRF3 operates at radio wavelengths, the implementation of the \textit{Gaia} mission \citep{2016A&A...595A...1G} led to the definition of the \textit{Gaia} Celestial Reference Frame 3 \citep[\textit{Gaia}-CRF3;][]{2022A&A...667A.148G} at optical wavelengths, extending high-precision CRF realizations beyond the radio domain.

The realization of a quasi-inertial CRF relies on the assumption that distant extragalactic radio sources are positionally fixed.
However, with the increasing number of VLBI observations and the improvement in astrometric precision, a number of radio sources have been found to exhibit positional instability and significant apparent motion \citep[e.g.,][]{1998AJ....116..516M, 2015AJ....150...58F, 2022MNRAS.512..874T, 2023AJ....165...69T, 2024AstL...50..657O}.
\citet{2018A&A...618A..80G} report that sources currently classified as positionally stable may reveal positional instabilities as observations accumulate.
Single-session source positions provide a direct way to investigate such variations, because they form coordinate time series for individual radio sources \citep[e.g.,][]{ma1997icrsdef, 2000A&A...359.1201F}.

Radio source positional variations may arise from several effects, including source structure evolution, brightness distribution changes, frequency-dependent core shifts, and observing geometry.
For example, the relationship between source astrophysical properties and positional stability has been investigated using indicators such as interstellar scintillation, flux-density variability, and color variability \citep[e.g.,][]{2013MNRAS.434..585S, 2014JGeod..88..575S, 2025A&A...695A.135L}.
\citet{2024A&A...684A..93L} reveal that higher photometric variabilities correspond to better positional stabilities and smaller optical-radio offsets, and a more stable CRF is realized through blazars \citep{2026A&A...706A.342S}.
Based on quad-band VLBI observations, \citet{2025AJ....169..173X} analyzed how source structure can lead to positional shifts.
Other studies have also discussed the impact of frequency-dependent core shifts \citep[e.g.,][]{2025MUPB...80S.143U} and observational geometry \citep[e.g.,][]{2018AstL...44..139T}.
These positional variations can affect the stability of the CRF and may propagate into geodetic products, such as EOPs and terrestrial reference-frame estimates \citep[e.g.,][]{2003JGRB..108.2275D,2005A&A...438.1141F, 2007JGeod..81..443M, 2007JGeod..81..455T, 2008A&A...481..535L, 2025A&AKrasna}.

Source coordinate time series are therefore useful for both reference-frame maintenance and astrophysical studies.
They have been used to evaluate ICRF stability \citep[e.g.,][]{2001A&A...375..661G, 2017MNRAS.466.1567L, 2022A&A...659A..75L}, select stable sources for future reference-frame realizations \citep[e.g.,][]{2003A&A...403..105F, 2009A&A...493..317L}, and model source coordinates as time-dependent parameters rather than constant catalog positions \citep{2017JGeod..91..755K}.
They also provide information on source-dependent astrometric behavior, such as the direction of positional variations relative to radio jets \citep[e.g.,][]{2011AJ....141..178M, 2021A&A...648A.125G, 2024ApJ...977L..14M}, periodic coordinate variations related to multiple-source systems \citep[e.g.,][]{2020A&A...634A.101R, 2024PASP..136e4503M, 2025A&A...700A.168G}, and large-scale apparent proper-motion fields associated with effects such as Galactic aberration, primordial gravitational waves, and anisotropic cosmic expansion \citep[e.g.,][]{2013A&A...559A..95T,2024A&A...688L..24L}.
A reliable assessment of the precision of radio source coordinate time series is therefore important for interpreting both source-dependent variations and their impact on the CRF.

Previous studies have investigated the noise properties of source coordinate time series using Allan standard deviation analysis \citep[e.g.,][]{2018A&A...618A..80G, 2024JPhCS2773a2007S}. 
\citet{2024ApJS..274...28C} assessed the quality of formal uncertainties by estimating excess noise from coordinate time series. 
Nevertheless, these studies mainly focused on individual time-series products.
Several International VLBI Service for Geodesy and Astrometry \citep[IVS;][]{Nothnagel2017} analysis centers provide source coordinate time-series solutions, which are produced using different data processing strategies and analysis configurations.
The consistency and intrinsic precision of these different products have not yet been systematically evaluated.
A comprehensive and independent assessment of the precision, reliability, and inter-solution consistency of radio source coordinate time series is therefore the primary motivation of this study.
 
In this work, we performed a comparative assessment of source coordinate time-series solutions from different IVS analysis centers.
We first applied data selection and reference-frame alignment, and then compared the solutions using common sources and common VLBI sessions.
Pairwise offsets between solutions were used to reduce common-mode signals and characterize inter-solution differences.
We further estimated the positional precision of each source in each solution using the N-cornered-hat (NCH) method \citep{2013A&A...558A..29M}.
Our goal is to quantify the external consistency and NCH-derived precision of radio source coordinate time series and investigate how different processing strategies, software packages, and parameter configurations affect the resulting time-series products.

The remainder of this paper is organized as follows. 
Section~\ref{sec:Data collection and preprocessing} describes the coordinate time-series products from different analysis centers and the data preprocessing procedure. 
The methods used to assess the precision among different coordinate time-series solutions are given in Sect.~\ref{sec:NCH method}.
Section~\ref{sec:Results} presents the coordinate time series of the example source 0923+392 and the main results of the precision analysis. 
Section~\ref{sec:Discussion} discusses the implications of the results, and Sect.~\ref{sec:Conclusion} summarizes the main conclusions.

\section{Data collection and preprocessing}
\label{sec:Data collection and preprocessing}
\subsection{Radio source coordinate time-series solutions}

\begin{table*}[ht!]
	\centering
    \caption{Summary statistics of the radio source coordinate time-series solutions provided by different analysis centers.}
	\begin{tabular}{lcccccc}
		\hline\hline
		         Solution & Nb. Src. & Nb. Sess. & Time span & Nb. Src. (filtered) & Processing mode & Software package \\
        (1) & (2) & (3) & (4) & (5) & (6) & (7) \\ \hline
        USNO             & 5385              & 7275               & 1980--2025 & 1161 & Global & Calc/Solve                         \\
        OPAs         & 5226              & 7435               & 1980--2025 & 1081 & Global & Calc/Solve                        \\
        OPAr         & 5817              & 7291               & 1980--2025 & 1148 & Independent & Calc/Solve                        \\
        AUS              & 5735              & 4927               & 1991--2025 & 1072 & Global & OCCAM                         \\
		OSO              & 5742              & 7478               & 1979--2025 & 875 & Independent & ASCOT                         \\
		VIE              & 5761              & 6962               & 1979--2025 & 1049 & Independent & VieVS                         \\
        UAVAC              & 4756              & 6598               & 1980--2023 & 741 & Independent & VieVS                         \\
        IAA         & 1911              & 6213               & 1979--2025 & 707 & Global & QUASAR                        \\ \hline
	\end{tabular}
    \tablefoot{The columns list: (1) solution identifier; (2) total number of sources; (3) total number of sessions; (4) time span; (5) number of sources after filtering; (6) processing mode; (7) software package used for the analysis.}
	\label{tab:number of sources}
\end{table*}

The radio source coordinate time-series solutions can be produced both in global and independent modes.
In global mode, sources are divided into $N$ subsets, each usually containing the same number of the ICRF3 defining and non-defining sources to preserve an identical and uniform sky distribution.
Subsequently, $N$-independent VLBI global solutions are conducted, in which positions of sources in each subset are treated as session-wise parameters to obtain their position time series.
Positions of the remaining sources are estimated as global parameters, among which the remaining ICRF3 defining sources are used to maintain the CRF in the global solution.
The complete time-series solutions for all sources can be derived through combining the $N$ solutions.
This method is denoted as an ``$N$-step'' solution in \citet{2023ivs..conf..298L}.
In the independent mode, the observables from each session are analyzed separately in the solution to derive local estimates of source positions, which usually suffers from a weakly constrained CRF.

We collected time-series solutions from multiple analysis centers, including United States Naval Observatory (USNO) IVS Analysis Center, Paris Observatory Geodetic VLBI Center (OPA), Geoscience Australia (AUS), Onsala Space Observatory (OSO), the Vienna IVS Analysis Center (VIE), Universidad de Alicante VLBI Analysis Center (UAVAC) and Institute of Applied Astronomy (IAA) of the Russian Academy of Sciences.
The strategies used to derive these time series differ among the analysis centers, leading to differences in the resulting products.
In this work, these differences were not treated merely as inconsistencies but were used to assess the external agreement among independent coordinate time series.
Such inter-solution comparisons provide an empirical means of evaluating the realistic precision of the source coordinate time series.

Table~\ref{tab:number of sources} summarizes the source coordinate time series collected from the different analysis centers, including the analysis strategy, time span, number of sessions, and number of sources for each dataset.
The USNO time-series solutions\footnote{\url{https://crf.usno.navy.mil/quarterly-vlbi-solution}} were produced using four-step global solutions with the Calc/Solve software \citep{1986AJ.....92.1020M}\footnote{\url{https://space-geodesy.nasa.gov/techniques/tools/calc_solve/calc_solve.html}}.
The USNO $S/X$-band dataset contains 5385 sources and covers the period from 1980 to 2025, including 7275 sessions.
The OPA provides solutions in two analysis configurations\footnote{\url{https://ivsopar.obspm.fr/radiosources/index.php}}.
The OPAs solutions were obtained from ten-step global solutions with Calc/Solve while the OPAr is the result of the operational solution, processing new observations as they become available to derive EOPs, source and station coordinates. It uses the independent mode of Calc/Solve. The OPAs solutions comprise 5226 sources based on 7435 sessions spanning 1980--2025, whereas the OPAr solution contains 5817 sources covering 7291 sessions over the same period. In the global solutions, all but four (0700--465, 0742--562, 0809--493, and 0918--534) of the 303 ICRF3 defining sources were used for the no-net-rotation (NNR) constraint. In OPAr, a loose NNR constraint was applied uniformly to all source coordinates.
The AUS solutions, produced at once using OCCAM~6.3 software \citep{2004ivsg.conf..267T} without imposing the NNR constraint, include 5735 sources and span 1991--2025, comprising a total of 4927 sessions.
The OSO solutions were derived from independent session-wise analyses using the ASCOT software \citep{2016ivs..conf..217A} and include 5742 sources over the period 1979--2025, comprising 7478 sessions.
The CRF was realized by the NNR constraint on the defining sources of ICRF3.
The VIE time-series solutions were generated by independent mode through two solutions \citep[$\text{vie}\_\text{sx}\_\text{250816a}$ and $\text{vie}\_\text{sx}\_\text{250816b}$;][]{krasna_2025} using the VieVS software \citep{2018PASP..130d4503B}\footnote{\url{http://vievswiki.geo.tuwien.ac.at/}}.
In the first solution, one half of the ICRF3 defining sources was fixed and therefore excluded from the time-series estimation, while in the second solution the other half was fixed.
In both solutions, the NNR constraint was not applied.
To obtain the full set of source time series, we combined these two solutions.
Sources present in the first solution but not in the second were added to the latter, resulting in a complete time-series dataset for 5761 sources from 1979 to 2025 with 6962 sessions.
The UAVAC time-series solutions were independently generated for each session using the VieVS software.
They include 4756 sources and 6598 sessions spanning 1980--2023.
The IAA time-series solutions were obtained from two-step global solutions \citep{2011AstL...37..267K}.
In the first solution, 295 ICRF2 defining sources were constrained, and time series for the non-defining sources were estimated.
In the second solution, 199 sources with longer observation histories and higher position accuracies were constrained, enabling the derivation of the time series of the defining sources.
The NNR constraint was applied to CRF on selected 203 defining sources.
These solutions were produced with the QUASAR software \citep{2007evga.conf.....B}, yielding coordinate time series for a total of 1911 sources from 1979 to 2025 with 6213 sessions.
These collected time-series solutions are used as the primary input data for this study and form the basis for the subsequent inter-solution comparisons and precision assessment.

\subsection{Data filtering}

Several criteria were adopted to reject outliers from the coordinate time-series solutions of each source.
First, for each source, data points derived from fewer than four observations in a given session were empirically excluded due to their lower reliability.
Second, for each session, we removed data points for which either the adjustment in right ascension or declination exceeded 50\,mas (to keep each coordinate consistent with its a priori position), formal error exceeded 30\,mas, or the correlation coefficient between right ascension and declination was equal to 1 or $-1$.
Third, if two or more solutions were available for the same session, only the latest one was retained.
Fourth, data points showing deviations smaller than 50\,mas in both right ascension and declination with respect to the two preceding and two following points in the time series were retained. This maintains the consistency and stability of time-series solutions.
Finally, sources with fewer than ten sessions were excluded.
The numbers of remaining sources are reported in Table~\ref{tab:number of sources}.
The loose thresholds introduced above filter outliers with extremely large uncertainties while preserving a relatively complete original dataset.

\subsection{Alignment to the ICRF3}

\begin{table*}[ht!]
	\centering
        \caption{Rotation and glide parameters between the mean source positions derived from different time-series solutions and the ICRF3 $S/X$ catalog.}
    \scalebox{1}{
	\begin{tabular}{lcccrrrrrr}
		\hline\hline
		         \multirow{2}{*}{Solution} & \multirow{2}{*}{Nb. Def. Src.} & \multirow{2}{*}{Nb. Used} & \multirow{2}{*}{Nb. Out} & $R_{1}$ & $R_{2}$ & $R_{3}$ & $G_{1}$ & $G_{2}$ & $G_{3}$ \\
                 &                &           &         & ($\mu$as) & ($\mu$as) & ($\mu$as) & ($\mu$as) & ($\mu$as) & ($\mu$as) \\
        (1) & (2)    & (3)    & (4)        & (5)     & (6)    & (7)     & (8)    & (9)     & (10)   \\ \hline
        USNO     & 297   & 295    & 2        & $+1\pm3$     & $+12\pm3$    & $+2\pm3$     & $+5\pm3$     & $+1\pm3$     & $+1\pm3$     \\
        OPAs & 262    & 259    & 3        & $+5\pm3$     & $+14\pm3$    & $+1\pm3$     & $+40\pm3$    & $-1\pm3$     & $+2\pm3$   \\
        OPAr & 299    & 210    & 89      & $-4\pm3$    & $-12\pm4$   & $-6\pm3$     & $-3\pm3$    & $+38\pm4$    & $+15\pm4$    \\
        AUS      & 293    & 287     & 6      & $-32\pm3$   & $+3\pm3$  & $-6\pm3$  & $+3\pm3$    & $-1\pm3$  & $-4\pm3$    \\
		OSO      & 299    & 285    & 14       & $-13\pm3$   & $-4\pm3$    & $-10\pm3$   & $+8\pm3$     & $-6\pm3$    & $-22\pm3$   \\
        VIE      & 299    & 296    & 3        & $+1\pm3$     & $-2\pm3$    & $-1\pm3$    & $-2\pm3$    & $-2\pm3$    & $-3\pm3$    \\
        UAVAC      & 260    & 247    & 13        & $-13\pm3$     & $-12\pm3$    & $+0\pm3$    & $+13\pm3$    & $-9\pm3$    & $-10\pm3$    \\
        IAA & 273    & 204    & 69        & $+4\pm4$     & $+9\pm4$    & $+27\pm4$     & $+8\pm4$    & $+8\pm4$     & $-2\pm4$   \\ \hline
	\end{tabular}}
    \tablefoot{The columns list: (1) solution identifier;
    (2) number of defining sources;
    (3) number of ICRF3 defining sources used in the estimation;
    (4) number of ICRF3 defining sources excluded from the estimation;
    (5)--(7) rotation parameters;
    and (8)--(10) glide parameters.
    All values are given as integers.}
	\label{tab:RG}
\end{table*}

To enable meaningful inter-solution comparisons, all time-series solutions were first aligned to ICRF3.
We constructed catalogs of weighted mean source positions from the time-series solutions.
These mean positions were estimated via the maximum likelihood (ML) method, as done in \citet{2024ApJS..274...28C}. 
The formulae are
\begin{equation}
\label{eq:ML_pos}
\begin{aligned}
& (\alpha_{ML},\delta_{ML})^T=\left( \sum_{i=1}^{n} C^{-1}_i\right)^{-1} {\sum_{i=1}^{n} C^{-1}_i}(\alpha_i,\delta_i)^T, \\
& C_i =\begin{pmatrix}
\sigma_{\alpha i}^2 & \rho_{i}\sigma_{\alpha i}\sigma_{\delta i}\\
\rho_{i}\sigma_{\alpha i}\sigma_{\delta i} & \sigma_{\delta i}^2
\end{pmatrix},
\end{aligned}
\end{equation}
where $\alpha_i$ and $\delta_i$ are the right ascension and declination at a single epoch, with formal errors $\sigma_{\alpha_i}$ and $\sigma_{\delta_i}$, respectively; $n$ is the number of data points; 
$\rho_{i}$ is the correlation between right ascension and declination; and $C_i$ is the corresponding covariance matrix.

The first degree of the vector spherical harmonics \citep[VSH;][]{2012A&A...547A..59M} technique was employed to investigate the differences between each weighted-mean catalog and ICRF3, following \citet{2022A&A...659A..75L}.
Specifically, \(\boldsymbol{R}=(R_{1}, R_{2}, R_{3})^{T}\) and \(\boldsymbol{G}=(G_{1}, G_{2}, G_{3})^{T}\) represent the rotation and the glide vectors, respectively. 
The rotation and glide parameters were determined by weighted least-squares fitting. 
The sources used to estimate these parameters are the ICRF3 defining sources available in each catalog.
Before estimating the rotation and glide parameters, sources with significant position deviations relative to their ICRF3 positions were removed using the ``555'' criterion \citep{2024AJ....167..229M}. 
That is, a source was identified as an outlier if the angular separation $D$ exceeded 5\,mas, the normalized separation $X$ exceeded 5, or the combined uncertainty of the source position difference exceeded 5\,mas. 
Approximately 90 sources from the OPAr catalogs were rejected by the ``555'' criterion, while nearly 70 sources were also excluded from the IAA catalog.
For the OPAr catalogs, the number of the ICRF3 defining sources remaining after this filtering is relatively small, which reduces the robustness of the estimated alignment parameters.

Table~\ref{tab:RG} presents the estimated rotation and glide parameters.
Most of the rotation parameters are smaller than $\mathrm{20\,\mu as}$, except for a few cases, such as $R_1$ of AUS reaching $\mathrm{-32\,\mu as}$, and $R_3$ of IAA reaching $\mathrm{27\,\mu as}$.
The glide parameters show more pronounced solution-dependent differences, with relatively large values of $G_1$ for OPAs, $G_2$ for OPAr, and $G_3$ for OSO.
Overall, the mean catalogs derived from the different time-series solutions are aligned with ICRF3 at the level of several tens of micro-arc seconds.

We also compared our results with those calculated in \citet{2024IERS.RG}, who analyzed VLBI global solutions from different analysis centers.
We find substantial differences between the two sets of rotation and glide parameters.
These differences may partly arise from the use of different source samples, since they used all available sources, whereas we used only the ICRF3 defining sources.
In addition, these two studies and our work are based on different types of source positions.
The catalog positions used in their studies were obtained from global solutions, in which a single position is estimated from all sessions, whereas our positions are weighted mean positions derived from coordinate time series.
The latter may be more sensitive to temporal sampling, data editing, and source-position variability.
We therefore applied a global rotation of $-\boldsymbol{R}$ to each session of the source coordinate time series for each solution separately.

\subsection{Final sample}
\label{Sec:final sample}
We first identified 583 sources common to all time-series datasets by source name, including 219 ICRF3 defining sources and 364 non-defining sources.
For each of these common sources, we extracted the data points corresponding to the same VLBI experiments (hereafter referred to as common sessions) from each time-series solution.
The time tags of the source positions for the same sessions were found to differ slightly among analysis centers.
Therefore, we adopted the time tags of USNO time series as the reference time tags.
After requiring at least ten common sessions for each source, the final sample contains 496 sources, including 195 ICRF3 defining sources and 301 non-defining sources.

\section{N-cornered-hat method}
\label{sec:NCH method}

We first formed pairwise offsets between time-series solutions for the same source and the same VLBI sessions.
In these paired differences, common-mode systematic errors are expected to be largely reduced, although solution-dependent systematics may still remain.
For each pair of time-series solutions, we then computed the weighted root mean square (WRMS) of the offsets in right ascension and declination separately.
The resulting WRMS values characterize the combined dispersion of each pair of solutions.
Finally, we applied the NCH method \citep[e.g.,][]{2007Handbook, 2013A&A...558A..29M} to separate these pairwise contributions and estimate the precision of each coordinate time-series solution for each source.

The WRMS is defined as
\begin{equation}
    \mathrm{WRMS}=\sqrt{\frac{\sum_{i=1}^{n}q_{i}(x_{i}-\bar{x})^2}{\sum_{i=1}^{n}q_{i}}},
    \label{Equ:WRMS}
\end{equation}
where $x_{i}$ is the offset in right ascension or declination between two solutions for the same session, $q_{i}$ is the weight given by the inverse squared uncertainty of $x_{i}$, $\bar{x}$ is the weighted mean of $x_{i}$, and $n$ is the number of common sessions.

For a set of $m$ time-series solutions, the classical NCH method is provided by \citet{2013A&A...558A..29M} as
\begin{equation}
\begin{aligned}
    & \sigma_{ab}^{2} = \sigma_{a}^{2} + \sigma_{b}^{2}, \\
    & \sigma_{a}^{2}  = \frac{1}{m-2}\left ( \sum_{b=1}^{m}\sigma_{ab}^{2}-B \right ), \\
    & B = \frac{1}{2(m-1)}\sum_{c=1}^{m}\sum_{b=1}^{m}\sigma_{cb}^{2}, \\
    & \sigma_{aa}=0,\sigma_{ab}=\sigma_{ba},
\end{aligned}
\end{equation}
where $ \sigma_{a}^{2}$ denotes the variance, or stochastic error level, of the $a$-th time-series solution and $ \sigma_{ab}^{2} $ represents the WRMS-derived variance of the paired offsets between the $a$-th and $b$-th solutions. 
This method may fail when it yields negative variances, which can occur if two paired time series are strongly correlated, exhibit widely significant discrepancies, or if the available data are insufficient.
Because the time-series solutions were derived from nearly identical input data with similar analysis strategies and software packages, it is necessary to apply the NCH method with correlations included:
\begin{equation}
    \sigma_{ab}^{2} = \sigma_{a}^{2} + \sigma_{b}^{2} - 2\rho_{ab}\sigma_{a}\sigma_{b},
\label{equ:nonlinear}
\end{equation}
where $\rho_{ab}$ is the correlation coefficient between the stochastic errors of the $a$-th and $b$-th solutions.
The estimation of reliable correlation coefficients is therefore essential.

We adopted an approximate method to estimate the correlations (denoted as $\rho_{ab}$) between two stochastic errors of two coordinate time-series solutions, similar to that described by \citet{2013A&A...558A..29M}.
First, for the same sessions, the offsets between solutions $a$ and $c$, as well as between solutions $b$ and $c$ were calculated and denoted by $\Delta_{ac}$ and $\Delta_{bc}$, where $c=1, ..., m$  with $c\neq a$ and $c\neq b$.
Second, the Pearson correlation coefficients $\rho_{ab}^{c}$ between two offsets series $\Delta_{ac}$ and $\Delta_{bc}$ were computed.
The general form of the weighted Pearson correlation coefficient is computed as
\begin{equation}
    \rho_{AB}=\frac{\sum_{i}^{n}\sqrt{\omega_{Ai}\omega_{Bi}}(A_{i}-\bar{A})(B_{i}-\bar{B})}{\sqrt{\sum_{i}^{n}\omega_{Ai}(A_{i}-\bar{A})^2\sum_{i}^{n}\omega_{Bi}(B_{i}-\bar{B})^2}},
\label{equ:correlation}
\end{equation}
where $\rho_{AB}$ is the weighted Pearson correlation coefficient between two series $A_{i}$ and $B_{i}$, with the same number of data points denoted as $n$; $\bar{A}$ and $\bar{B}$ are the weighted mean computed by the ML method following Eq.~(\ref{eq:ML_pos}); and $\omega_{Ai}$ and $\omega_{Bi}$ are the weights given by the inverse of squared uncertainties.
Third, the average of $\rho_{ab}^{c}$ over all possible choices of $c$ was taken as the estimation of the correlation between stochastic errors of coordinate time-series solutions, $a$ and $b$.

Based on the above approaches, we performed bootstrap resampling \citep{1979AS...7...1} 1000 times for each source.
For each bootstrap realization of a given source, we first generated a sequence of indices via a random number generator using \texttt{random.default\_rng} function in Python \texttt{NumPy} library with fixed random seeds to ensure reproducibility.
The same set of indices was then applied to all time-series solutions, so that the paired common sessions were preserved in the resampled dataset.
For each resampled dataset, we estimated the correlation coefficients, applied the NCH method and solved Eq.~(\ref{equ:nonlinear}) via nonlinear least-squares fitting using the \texttt{optimize.least\_squares} function in the Python \texttt{SciPy} library.
The final NCH estimates were taken as the medians of all bootstrap realizations, and their uncertainties were derived from the 16th and 84th percentiles, corresponding to the $68\%$ central confidence interval.
The NCH analysis was carried out separately for right ascension and declination, yielding stochastic error estimates for each source in each time-series solution.

\section{Results}
\label{sec:Results}
\subsection{Comparison of source coordinate time-series solutions}

\begin{figure*}[htbp]
  \centering
  \includegraphics[width=\textwidth]{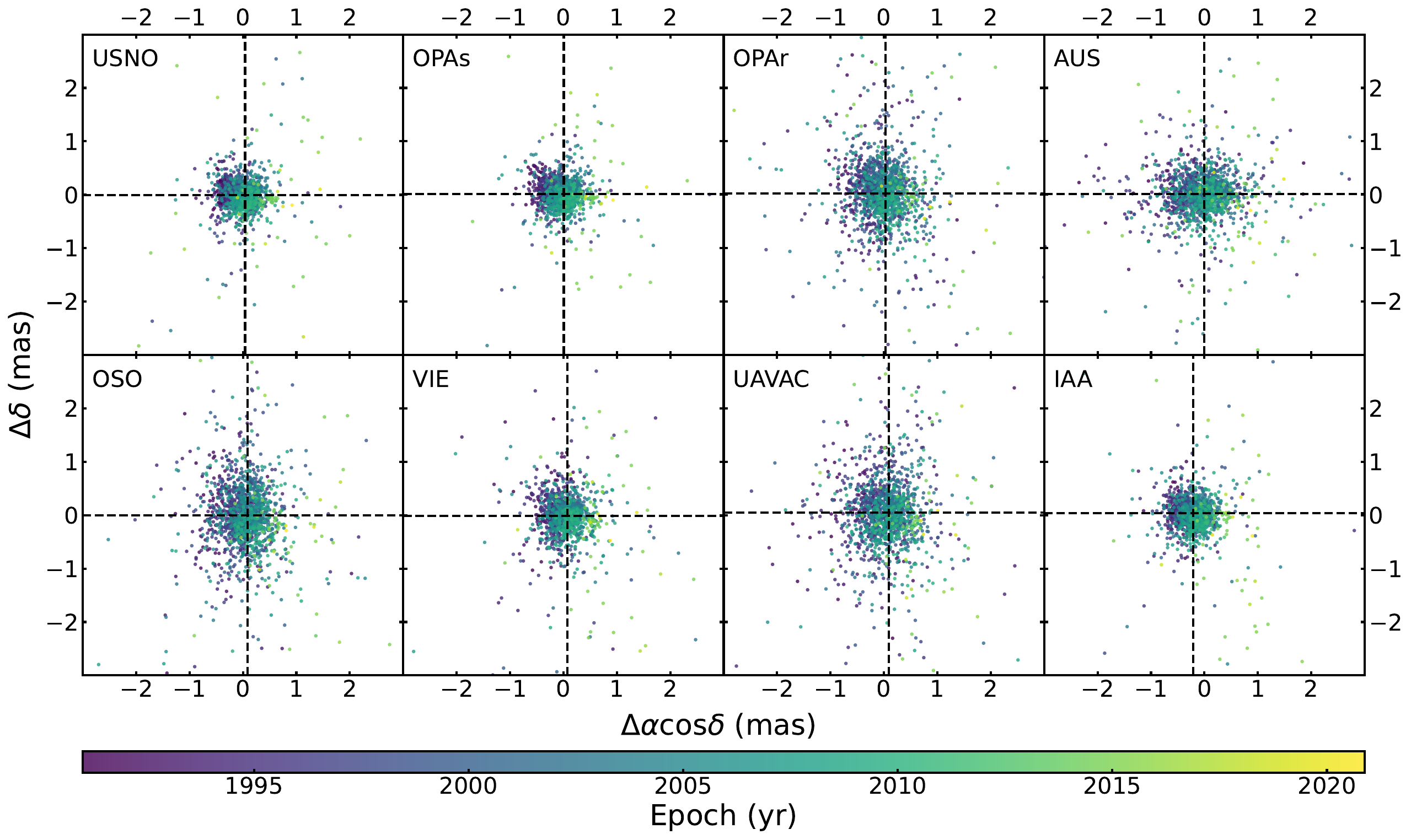}
  \caption{Coordinate time-series solutions for source 0923+392. 
    Each panel corresponds to one solution, labeled in the top left corner.
    The origin corresponds to the ICRF3 position. 
    The plotted ranges in right ascension and declination range from $-3$ to $3$ mas. 
    The dashed lines indicate the ML positions, and the color bar gives the session epochs.}
  \label{fig:combine offsets time series}
\end{figure*}

\begin{figure*}[hbtp]
    \centering
    \includegraphics[width=0.49\textwidth]{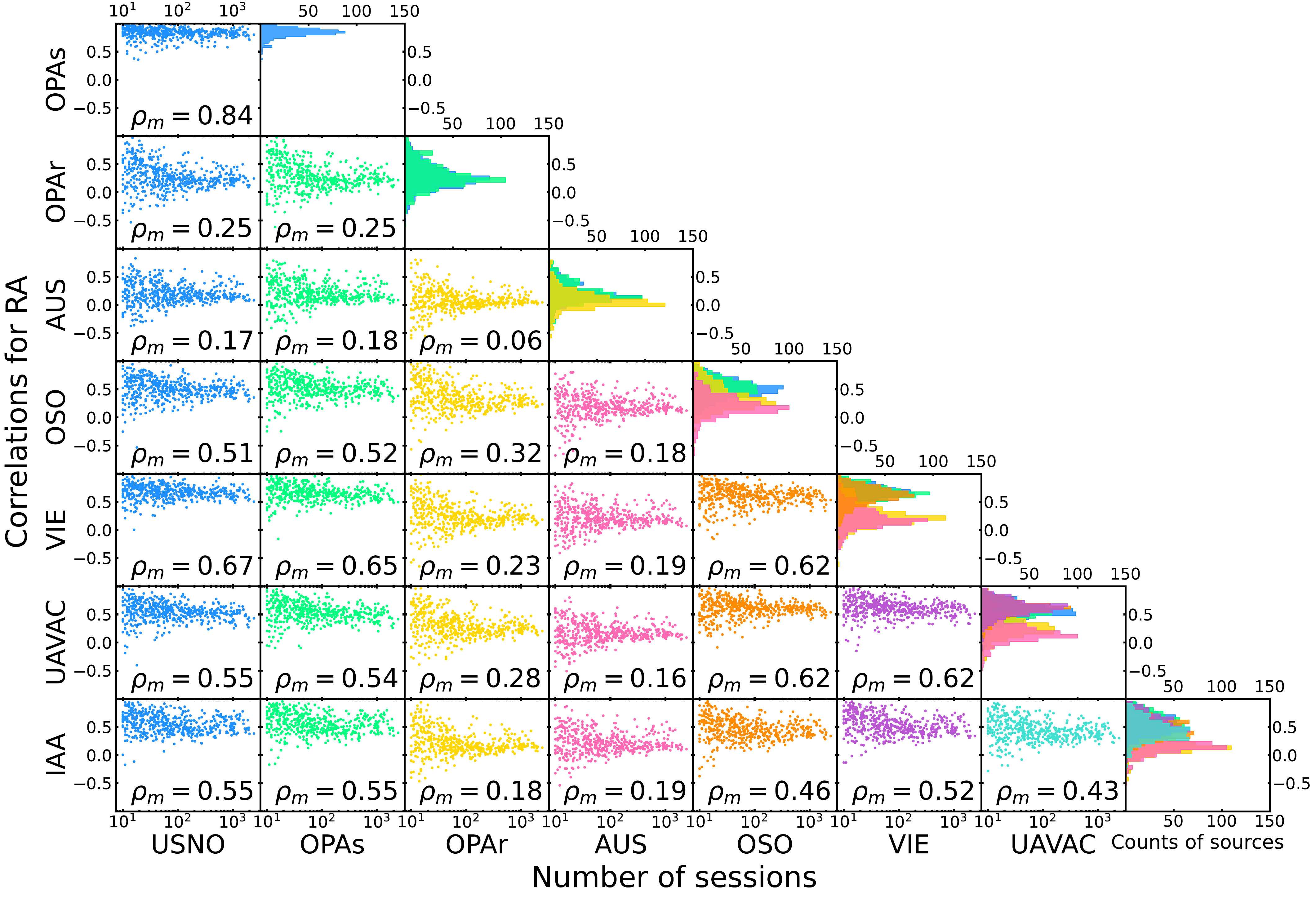}
    \includegraphics[width=0.49\textwidth]{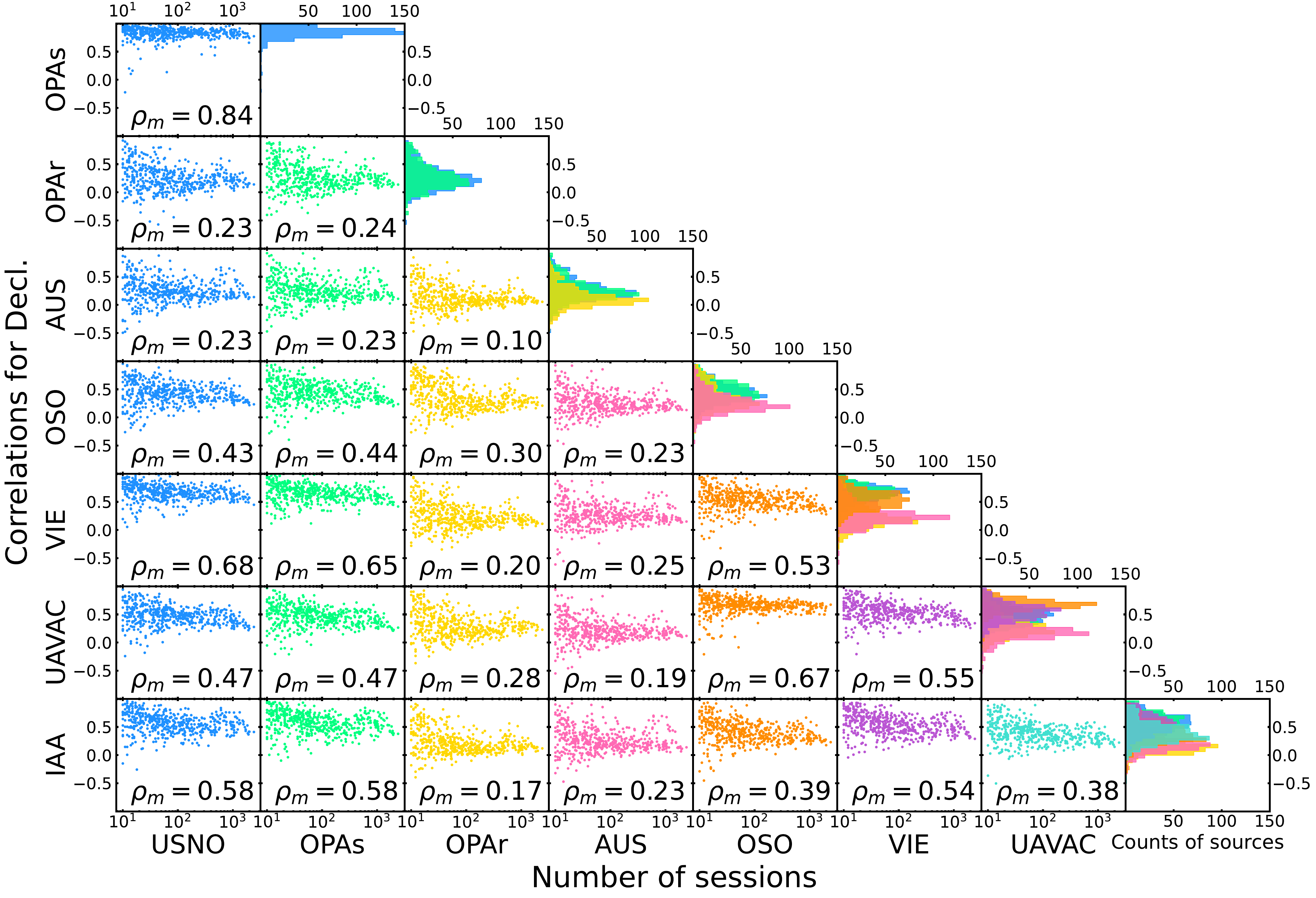}
    \caption{Pearson correlation coefficients as a function of the number of sessions for 496 sources in the final sample between each pair of coordinate time-series solutions, shown separately for right ascension (left) and declination (right). 
    The histograms on the right, with 20 bins, show the distributions of the Pearson correlation coefficients.
    The median correlation for each pair is given in the bottom right corner of each subplot.}
    \label{figA:correlations}
\end{figure*}

\begin{figure*}[hbtp]
    \centering
    \includegraphics[width=0.49\textwidth]{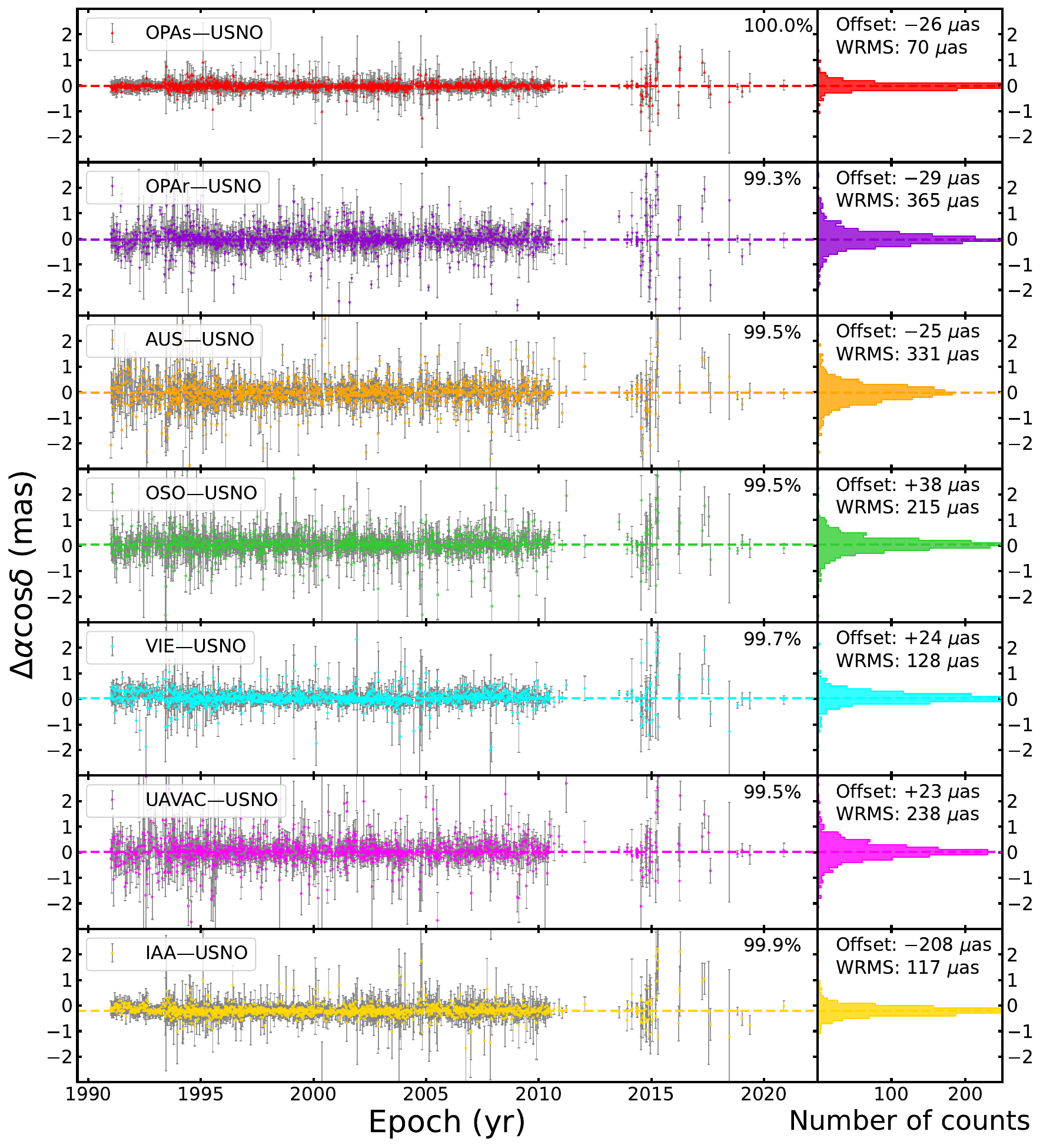}
    \includegraphics[width=0.49\textwidth]{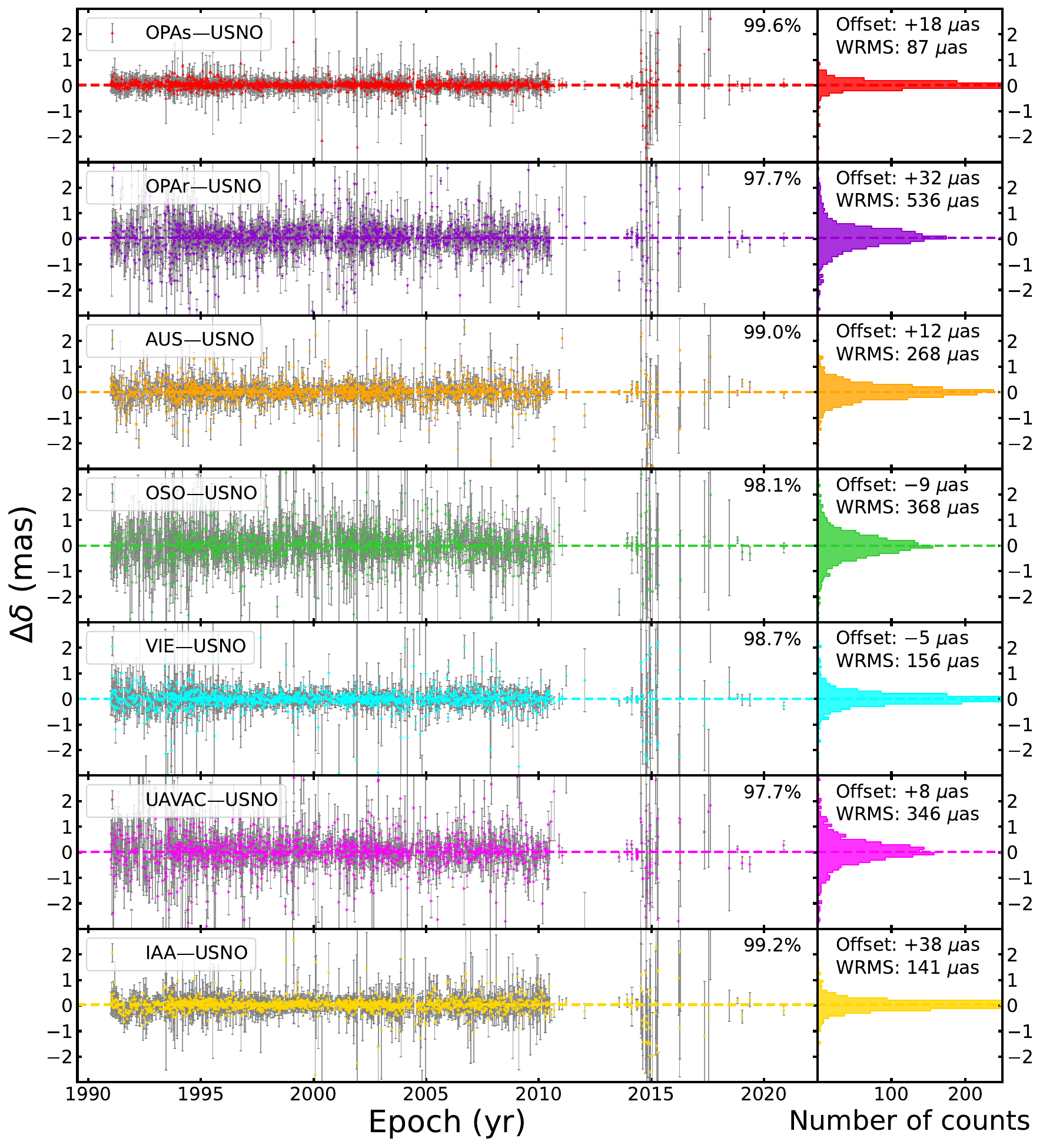}
    \caption{Positional offsets of seven time-series solutions relative to the USNO solution for the corresponding sessions of source 0923+392, shown separately for right ascension (left) and declination (right).
    The reference position for each session corresponds to the USNO time-series solution.
    The mean offset and WRMS for each offset time series, rounded to the nearest integer, are displayed in the top right corner of the corresponding histogram.}
    \label{figA:same session offset from USNO for 0923+392}
\end{figure*}

\begin{table*}[ht]
	\centering
    \caption{Statistics of the ML position offsets and WRMS values of the source coordinate time-series solutions with respect to their ICRF3 position at the $S/X$ band for 496 sources in the final sample.}
	\begin{tabular}{lcccccccc}
		\hline \hline
        \multirow{3}{*}{Solution} & \multicolumn{2}{c}{RA offsets} & \multicolumn{2}{c}{Decl. offsets} & \multicolumn{2}{c}{RA WRMS} & \multicolumn{2}{c}{Decl. WRMS} \\
        & \multicolumn{2}{c}{($\mu$as)}   & \multicolumn{2}{c}{($\mu$as)} & \multicolumn{2}{c}{($\mu$as)}   & \multicolumn{2}{c}{($\mu$as)} \\
        \cline{2-9}
		& Range               & Median              & Range              & Median             & Range              & Median             & Range             & Median            \\
		(1)                              & (2)                 & \multicolumn{1}{c}{(3)}                 & (4)                & \multicolumn{1}{c}{(5)}                & (6)                & (7)                & (8)               & (9)               \\ \hline
		USNO                             & [$-484$, $+705$]        & $+2$                   & [$-611$, $+389$]       & $+4$                  & [$59$, $1633$]       & $203$                  & [$93$, $3090$]      & $238$                 \\
		OPAs                             & [$-271$, $+831$]        & $+3$                   & [$-404$, $+662$]       & $+5$                  & [$75$, $4430$]       & $204$                  & [$69$, $3390$]      & $240$                 \\
		OPAr                             & [$-890$, $+1314$]        & $+0$                   & [$-3976$, $+1902$]       & $+48$                  & [$174$, $4973$]       & $697$                  & [$151$, $8101$]      & $913$                 \\
		AUS                              & [$-696$, $+920$]        & $-1$                   & [$-1347$, $+925$]       & $+1$                  & [$137$, $3861$]       & $302$                  & [$118$, $4461$]      & $237$                 \\
		OSO                              & [$-383$, $+836$]        & $+8$                   & [$-671$, $+544$]       & $-19$                  & [$110$, $1924$]       & $296$                  & [$128$, $2781$]      & $437$                 \\
		VIE                              & [$-644$, $+635$]        & $+3$                   & [$-555$, $+517$]       & $-5$                  & [$86$, $1953$]       & $227$                  & [$117$, $2686$]      & $272$                 \\
		UAVAC                            & [$-507$, $+660$]        & $+0$                   & [$-989$, $+695$]       & $+2$                  & [$147$, $1638$]       & $292$                  & [$152$, $2743$]      & $425$                 \\
		IAA                              & [$-6356$, $+7443$]        & $-15$                   & [$-37592$, $+7797$]       & $+3$                  & [$86$, $3553$]       & $289$                  & [$106$, $2655$]      & $324$                 \\ \hline
    	\end{tabular}
            \tablefoot{The columns list the following: 
            (1) solution identifier;
            (2), (3) range and median of the offsets from the ICRF3 positions in right ascension;
            (4), (5) range and median of the offsets from the ICRF3 positions in declination;
            (6), (7) range and median of the WRMS values in right ascension; 
            and (8), (9) range and median of the WRMS values in declination.
            All values are rounded to the nearest integer.}
	\label{tabA:offsets and WRMS for all sources}
\end{table*}

For each source in the final sample, we computed the coordinate offsets of all time-series solutions with respect to the positions provided by the ICRF3 catalog and computed the corresponding WRMS values to quantify the scatter of these offsets in right ascension and declination.
We used source 0923+392 (4C 39.25) as an example to illustrate the differences among the time-series solutions.
This source was selected as a typical example owing to its long observational history and obvious nonlinear variations in the coordinate time series, which validates the effective mitigation of systematic effects in the subsequent analysis.
We also investigated two sources with highly stable positions, namely 0059+581 and 0851+202 (OJ 287).
Their analytical results present similar features to those of source 0923+392.

For 0923+392, all time-series solutions remain tightly clustered with respect to the ICRF3 position, indicating relatively small source position variations, as shown in Fig.~\ref{fig:combine offsets time series}.
All solutions exhibit a similar nonlinear position variation from northwest to southeast, despite the small offsets presented among the solutions in the individual session-wise estimates.
The ML positions of all solutions differ only slightly from the ICRF3 position, generally at the level of a few micro-arc seconds, although somewhat larger offsets in right ascension are found for the OSO $(+86\,\mathrm{\mu as})$, VIE $(+74\,\mathrm{\mu as})$, UAVAC $(+92\,\mathrm{\mu as})$, and IAA $(-207\,\mathrm{\mu as})$ solutions, where the IAA solution shows the largest deviation.
All solutions show approximately circular scatter distributions, while the OPAr, AUS, OSO, and UAVAC solutions exhibit greater scatter than the others.
A more detailed quantification of the offsets and scatter in right ascension and declination, as well as the temporal variations of source 0923+392, is provided in Appendix~\ref{appendix:coordinate_vs_t}.

The statistics of offsets and WRMS for all sources from different coordinate time-series solutions are presented in Table~\ref{tabA:offsets and WRMS for all sources}, while the corresponding distributions are shown in Appendix~\ref{appendix:coordinate_vs_t}.
For most solutions, the median offsets in both right ascension and declination are close to zero, generally at the level of several micro-arc seconds.
The main exceptions are the right ascension offset of IAA, with a median of $-15\,\mathrm{\mu as}$, the declination offset of OPAr, with a median of $+48\,\mathrm{\mu as}$, and the declination offset of OSO, with a median of $-19\,\mathrm{\mu as}$.
The median WRMS values show clearer differences among the solutions.
In right ascension, USNO, OPAs, and VIE have median WRMS values of about $200$--$230\,\mathrm{\mu as}$, whereas AUS, OSO, UAVAC, and IAA have slightly larger values of about $290$--$300\,\mathrm{\mu as}$.
In declination, USNO, OPAs, AUS, and VIE show median WRMS values of about $240$--$270\,\mathrm{\mu as}$, while OSO, UAVAC, and IAA have larger values.
The OPAr solution is the most scattered in both coordinates, with median WRMS values of $697\,\mathrm{\mu as}$ in right ascension and $913\,\mathrm{\mu as}$ in declination.
The IAA solution shows the largest offset ranges in both right ascension and declination, especially in declination.
However, its median WRMS values remain comparable to those of most other solutions, except OPAr.
This suggests that the large offset range of IAA is mainly caused by a subset of sources with large mean-position offsets relative to ICRF3 rather than by a generally larger session-wise scatter of the time series.

To further assess the consistency between different time-series solutions, we calculated the Pearson correlation coefficients between each pair of solutions for each source in the final sample.
The coefficients were computed separately for right ascension and declination using Eq.~(\ref{equ:correlation}), and the results for 0923+392 are illustrated in Appendix~\ref{secA:Consistency}.
We also tested the effect of removing long-term linear trends from the time series before calculating the correlation coefficients using the method described in Sect.~\ref{sec:NCH method}, and find that the results changed only marginally.
Therefore, the following discussion is based on the coordinate offset time series without detrending.
The correlations in right ascension and declination for all time-series solution pairs across all sources are presented in Fig.~\ref{figA:correlations}.
For both right ascension and declination, the correlation coefficients become more stable for sources with a larger number of common sessions, although their median values differ substantially among the solution pairs.
In right ascension, the USNO--OPAs pair converges to the highest median correlation of $0.84$, whereas the OPAr--AUS pair converges to the median lowest correlation of $0.06$.
A similar pattern is also seen in declination: the USNO--OPAs pair shows the highest median correlation of $0.84$, while the OPAr--AUS pair shows the lowest median correlation of $0.10$.

As mentioned earlier, coordinate time-series solutions may be affected by common-mode systematic errors.
By forming positional offsets between different solutions for the same sessions, these common components are expected to be largely reduced, thereby allowing the solution-dependent differences to be examined more clearly.
As an example, Fig.~\ref{figA:same session offset from USNO for 0923+392} shows the offset time series for 0923+392, with the USNO solution taken as the reference for each session. 
For this source, the nonlinear positional variations in right ascension and declination are largely removed.
Most of the offsets remain within $\pm\,1\,\mathrm{mas}$, indicating good consistency between the USNO solution and the other time-series solutions. 
A notable exception is the IAA solution, which shows a mean right ascension offset of $-208\,\mathrm{\mu as}$ relative to USNO.
The WRMS values provide a quantitative measure of the scatter of these pairwise offset time series.
In right ascension, the WRMS ranges from $70\,\mathrm{\mu as}$ for OPAs relative to USNO to $365\,\mathrm{\mu as}$ for OPAr relative to USNO.
In declination, the corresponding WRMS ranges from $87\,\mathrm{\mu as}$ for OPAs relative to USNO to $536\,\mathrm{\mu as}$ for OPAr relative to USNO.
The larger WRMS values of OPAr, AUS, OSO, and UAVAC indicate larger session-wise positional differences with respect to the USNO reference.

\subsection{N-cornered-hat-derived positional precision of time-series solutions}
\label{subsec:NCH_results}

\begin{table*}[ht]
	\centering
    \caption{Statistics of NCH-derived precision and formal errors for different source coordinate time-series solutions for 481 sources in right ascension and for 474 sources in declination.}
	\begin{tabular}{lcccccccc}
		\hline \hline
		\multirow{3}{*}{Solution} & \multicolumn{2}{c}{Precision for RA}    & \multicolumn{2}{c}{Precision for Decl.}    & \multicolumn{2}{c}{Formal error for RA}     & \multicolumn{2}{c}{Formal error for Decl.}     \\
        & \multicolumn{2}{c}{($\mu$as)}   & \multicolumn{2}{c}{($\mu$as)} & \multicolumn{2}{c}{($\mu$as)}   & \multicolumn{2}{c}{($\mu$as)} \\
        \cline{2-9}
		                                 & Range               & Median              & Range              & Median             & Range              & Median             & Range             & Median            \\
		(1)                              & (2)                 & \multicolumn{1}{c}{(3)}                 & (4)                & \multicolumn{1}{c}{(5)}                & (6)                & (7)                & (8)               & (9)               \\ \hline
		USNO                             & [$16$, $847$]       & $213$                  & [$10$, $863$]      & $263$                  & [$78$, $1993$]        & $213$                  & [$70$, $1361$]       & $249$          \\
		OPAs                             & [$2$, $2494$]       & $213$                  & [$56$, $844$]      & $259$                 & [$74$, $2211$]        & $201$                   & [$67$, $1344$]       & $238$   \\
		OPAr                             & [$139$, $4770$]       & $626$                  & [$162$, $3196$]      & $824$                 & [$144$, $2265$]        & $273$                   & [$150$, $1378$]       & $354$    \\
		AUS                              & [$21$, $1575$]       & $407$                  & [$130$, $1190$]      & $355$                 & [$97$, $1111$]        & $256$                   & [$90$, $899$]       & $204$      \\
		OSO                              & [$32$, $1423$]       & $266$                  & [$14$, $1696$]      & $398$                 & [$120$, $1705$]        & $263$                   & [$131$, $1936$]       & $377$    \\
		VIE                              & [$29$, $1087$]       & $210$                  & [$30$, $923$]      & $270$                 & [$77$, $1230$]        & $206$                   & [$74$, $1365$]       & $248$     \\
		UAVAC                            & [$20$, $1705$]       & $272$                  & [$93$, $1417$]      & $395$                 & [$128$, $1991$]        & $242$                   & [$122$, $1305$]       & $335$      \\
		IAA                              & [$2$, $1729$]       & $265$                  & [$20$, $1393$]      & $310$                 & [$145$, $3354$]        & $371$                   & [$143$, $2400$]       & $433$    \\ \hline
	\end{tabular}
    \tablefoot{The columns list the following:  
    (1) solution identifier;
    (2), (3) range and median of the NCH-derived precision in right ascension;
    (4), (5) range and median of the NCH-derived precision in declination;
    (6), (7) range and median of the formal errors in right ascension;
    (8), (9) range and median of the formal errors in declination.
    All values are rounded to the nearest integer.}
	\label{tab:formal error and accuracy}
\end{table*}

The positional precision of the coordinate time-series solutions was estimated through the method described in Sect.~\ref{sec:NCH method}.
After 1000 bootstrap realizations, positive median NCH estimates were obtained for 481 sources in right ascension and 474 sources in declination.

For the example source 0923+392, the NCH-derived precision shows clear differences among the time-series solutions.
In right ascension, the precision ranges from $133\,\mathrm{\mu as}$ for VIE to $363\,\mathrm{\mu as}$ for OPAr, while in declination it ranges from $161\,\mathrm{\mu as}$ for VIE to $542\,\mathrm{\mu as}$ for OPAr.
The USNO and OPAs solutions give very similar precision estimates, with values of $166\,\mathrm{\mu as}$ and $163\,\mathrm{\mu as}$ in right ascension, and $210\,\mathrm{\mu as}$ and $201\,\mathrm{\mu as}$ in declination, respectively.
The OPAr solution shows the poorest precision for this source in both coordinates, consistent with its larger WRMS values in the pairwise offset time series discussed above.
A more detailed comparison of the formal-error time series for 0923+392 is provided in Appendix~\ref{secA:Comparison between precision and formal errors}.

Table~\ref{tab:formal error and accuracy} presents the source-wise statistics of the NCH-derived precision estimates and the formal errors for all sources with positive median NCH estimates.
For each source and each solution, the NCH-derived precision is obtained from the NCH analysis, whereas the formal uncertainty is represented by the median formal error over all common sessions selected in Sect.~\ref{Sec:final sample}.

In right ascension, the median NCH-derived precision is about $210\,\mathrm{\mu as}$ for USNO, OPAs, and VIE, indicating a similar level of positional precision among these three solutions.
The median values are slightly larger for OSO, UAVAC, and IAA, ranging from $265$ to $272\,\mathrm{\mu as}$.
The AUS solution shows a larger value of $407\,\mathrm{\mu as}$, while OPAr has the largest median value, reaching $626\,\mathrm{\mu as}$.

In declination, the median NCH-derived precision is generally poorer than that in right ascension, except for the AUS solution.
For the AUS solution, the median NCH-derived precision estimate is greater in right ascension than in declination, corresponding to a poorer precision in right ascension ($407\,\mathrm{\mu as}$) than in declination ($355\,\mathrm{\mu as}$).
The USNO, OPAs, and VIE solutions again show comparable values, with medians of about $260$--$270\,\mathrm{\mu as}$.
The IAA solution has an intermediate median value of $310\,\mathrm{\mu as}$, while the OSO and UAVAC solutions show larger values close to $400\,\mathrm{\mu as}$.
The OPAr solution again gives the poorest result, with a median precision of $824\,\mathrm{\mu as}$.

The formal errors show a different pattern.
In right ascension, the median formal errors of most solutions lie between $200$ and $270\,\mathrm{\mu as}$, except for IAA, which has the largest median formal error of $371\,\mathrm{\mu as}$.
In declination, the median formal errors range from $204\,\mathrm{\mu as}$ for AUS to $433\,\mathrm{\mu as}$ for IAA.
Overall, except for IAA in both coordinates and USNO in right ascension, the median values of NCH-derived precision estimates are larger than the corresponding median formal errors.
This suggests that the formal errors generally underestimate the external scatter revealed by the inter-solution comparison.
The differences are relatively small for USNO, OPAs, and VIE, indicating broad consistency between their NCH-derived precision estimates and formal errors.
In contrast, the OPAr solution shows NCH-derived precision estimates much larger than its formal errors in both coordinates, suggesting that its formal errors do not fully reflect the external scatter.
For IAA, the median formal errors are larger than the NCH-derived precision estimates in both coordinates, indicating that its formal errors may be conservative.
A more detailed comparison between the NCH-derived precision estimates and the formal errors is
discussed in Sect.~\ref{sec:Precision and formal error}.

We examined variations in the NCH-derived positional precision of individual sources as a function of the number of observational sessions, right ascension and declination, and fit a smoothed trend to these dependencies.
The smoothed trend is obtained by employing the Nadaraya-Watson kernel regression approach \citep{1964TPA...9...141, 1964...26...359} using the \texttt{Kernelreg} function in the \texttt{statsmodels} library.
The detailed dependence of the NCH-derived source positional precision on the number of observational sessions is presented in Fig.~\ref{figA:accuracy scatter sessions}.
These results suggest that a larger number of sessions does not necessarily lead to smaller NCH-derived precision estimates.
Instead, the typical precision level appears to be largely solution-dependent, suggesting the presence of a solution-specific precision floor.
Nevertheless, the NCH estimates are more stable for sources with more observational data.
In contrast to the clear dependence on declination in Fig.~\ref{figA:accuracy scatter declination}, the precision estimates show no strong systematic dependence on right ascension, with only weak nonlinear variations for some solutions in Fig.~\ref{figA:accuracy scatter right ascension}.

Some extremely small precision values, $2\,\mathrm{\mu as}$, are found for the OPAs and IAA solutions in right ascension.
The smallest value for OPAs occurs for 0537--286, with the 16th and 84th percentiles of $-621\,\mathrm{\mu as}$ and $401\,\mathrm{\mu as}$, respectively.
The smallest value for IAA occurs for 0823--223, with the corresponding percentiles of $-130\,\mathrm{\mu as}$ and $343\,\mathrm{\mu as}$.
These broad percentile ranges, together with the negative lower bounds, indicate that the NCH estimates for these sources are poorly constrained.

We inspected the time series of these two sources to identify the possible causes.
For 0537--286, although 225 data points are available, the pairwise offset time series show very large inter-solution scatter, with the WRMS of the OPAr--USNO pair reaching $1672\,\mathrm{\mu as}$.
For 0823--223, only 11 observational sessions are available, and the small number of data points likely leads to an unstable NCH estimate.
These examples illustrate that the NCH-derived precision estimates can become unreliable when the pairwise differences are highly dispersed or when the number of common sessions is insufficient.
A more general assessment of the limitations of the NCH method is given in Sect.~\ref{sect:NCH-shortcoming}.

\section{Discussion}
\label{sec:Discussion}
\subsection{Impact of analysis strategy on coordinate time series}

\begin{table*}[htbp]
	\centering
    \caption{Median absolute differences between source-wise WRMS values for each pair of coordinate time-series solutions for 496 sources in the final sample.}
	\begin{tabular}{lccccccc}
		\hline \hline
		  \multirow{2}{*}{Solution}    & USNO    & OPAs    & OPAr    & AUS    & OSO    & VIE    & UAVAC \\
         & ($\mu$as) & ($\mu$as) & ($\mu$as) & ($\mu$as) & ($\mu$as) & ($\mu$as) & ($\mu$as)\\
        \hline
		OPAs  & 11, 16   &         &         &        &        &        &       \\
		OPAr  & 466, 644 & 460, 631 &         &        &        &        &       \\
		AUS   & 104, 49  & 101, 48  & 378, 664 &        &        &        &       \\
		OSO   & 90, 178  & 87, 179  & 365, 434 & 63, 179 &        &        &       \\
		VIE   & 30, 35   & 27, 36   & 440, 598 & 85, 49  & 65, 147 &        &       \\
		UAVAC & 91, 169  & 91, 168  & 375, 454 & 60, 172 & 28, 40  & 66, 142 &       \\
		IAA   & 80, 81   & 80, 80   & 388, 536 & 64, 96  & 52, 104 & 61, 61  & 52, 94 \\ \hline
	\end{tabular}
    \tablefoot{All values are given in the order of right ascension and declination, rounded to the nearest integer.}
	\label{tab:WRMS compare across solutions}
\end{table*}

\begin{table*}[htbp]
	\centering
    \caption{Median absolute differences between source-wise NCH-derived precision estimates for each pair of coordinate time-series solutions for 481 sources in right ascension and for 474 sources in declination.}
	\begin{tabular}{lccccccc}
		\hline \hline
		  \multirow{2}{*}{Solution}    & USNO    & OPAs    & OPAr    & AUS    & OSO    & VIE    & UAVAC \\
        & ($\mu$as) & ($\mu$as) & ($\mu$as) & ($\mu$as) & ($\mu$as) & ($\mu$as) & ($\mu$as)\\
        \hline
		OPAs  & 14, 22   &         &         &        &        &        &       \\
		OPAr  & 386, 518 & 388, 523 &         &        &        &        &       \\
		AUS   & 194, 102  & 194, 107  & 199, 416 &        &        &        &       \\
		OSO   & 64, 133  & 61, 140  & 313, 378 & 132, 64 &        &        &       \\
		VIE   & 29, 37   & 26, 37   & 384, 522 & 193, 103  & 62, 133 &        &       \\
		UAVAC & 64, 138  & 63, 139  & 331, 383 & 133, 63 & 35, 38  & 65, 132 &       \\
		IAA   & 72, 91   & 75, 88   & 318, 452 & 135, 75  & 68, 109 & 79, 89  & 59, 115 \\ \hline
	\end{tabular}
    \tablefoot{All quantities are listed in order of right ascension and declination, rounded to the nearest integer.}
	\label{tab:precision compare across solutions}
\end{table*}

Although the coordinate time-series solutions are derived from largely the same VLBI observations, they are not identical products.
Differences in processing strategy, software implementation, parameterization, and source constraints can all affect the resulting source coordinates.
Therefore, comparisons of the offsets, WRMS values, correlations, and NCH-derived precision estimates provide a useful way to assess how analysis strategies influence the external consistency of the coordinate time series.

Two broad processing strategies are involved in the time-series solutions analyzed in this work: global-mode and independent session-wise solutions.
The global-mode solutions generally estimate source positions in each session with a global adjustment to the positions of other sources that maintain the CRF, whereas the independent-mode solutions estimate source positions separately for individual sessions.
However, this distinction alone does not fully describe the differences among the products, because the detailed choices of global and local parameters, source constraints, and software implementations also differ among analysis centers.

The mean-position offsets with respect to the ICRF3 positions provide the first indication of such strategy-dependent behavior.
As shown in Table~\ref{tabA:offsets and WRMS for all sources}, large offset ranges are found for the IAA solution in both right ascension and declination.
This may be related to its two-step global strategy, in which the defining sources of ICRF2 and a selected set of well-observed sources are used to derive the time series for other sources.
Such a configuration differs from the strategies adopted by the other global solutions and suggests that the choice of global and local parameters can have a substantial effect on the derived mean source positions.
For the OPAr solution, the relatively large offsets and the large number of defining sources rejected in the alignment procedure are likely related to the larger session-wise scatter of its independent solution, as also reflected by its large WRMS values.

The WRMS comparison further provides a descriptive measure of how similar the scatter levels are between different solutions. 
For each pair of solutions, we calculated the absolute difference between their WRMS values for the same sources and adopted the median over all sources as a summary statistic, as listed in Table~\ref{tab:WRMS compare across solutions}.
Although this quantity does not represent a formal statistical test, it is useful for identifying pairs of solutions with similar source-wise scatter levels.

The smallest differences are found for the USNO--OPAs pair, with median absolute WRMS differences of only $11\,\mathrm{\mu as}$ in right ascension and $16\,\mathrm{\mu as}$ in declination.
Small differences are also found for the USNO--VIE and OPAs--VIE pairs, with values of $30$ and $35\,\mathrm{\mu as}$ in right ascension and declination for the USNO--VIE pair, and $27$ and $36\,\mathrm{\mu as}$ for the OPAs--VIE pair.
These results indicate that the WRMS levels of USNO, OPAs, and VIE are highly similar for the same sources.
The strong agreement between USNO and OPAs is expected, because both solutions were produced in the global mode using the Calc/Solve software.
The similarity of VIE to these two solutions is likely related to its processing strategy, in which subsets of ICRF3 defining sources were fixed in two complementary session-wise solutions.
The fixing of chosen defining sources enables a stable datum definition with respect to ICRF3.
A limitation of this approach is that possible uncertainties in one or more of the fixed components directly affect all other components to be estimated.

Among the other independent session-wise solutions, the OSO--UAVAC pair shows the closest agreement, with median absolute WRMS differences of $28\,\mathrm{\mu as}$ in right ascension and $40\,\mathrm{\mu as}$ in declination.
This suggests that these two solutions have similar scatter levels, despite being produced with different software packages.
In contrast, the OPAr solution shows much larger differences with all other solutions, typically exceeding $360\,\mathrm{\mu as}$ in right ascension and $430\,\mathrm{\mu as}$ in declination.
This confirms that OPAr has a substantially larger source-wise scatter level than the other time-series solutions.

The correlation coefficients between paired coordinate time series further support this interpretation.
In general, solutions with more similar processing methods tend to show stronger correlations.
The strongest correlations are found between the USNO and OPAs solutions, which share both the global-mode strategy and the Calc/Solve software.
In contrast, the weakest correlations are found between OPAr and AUS, which differ in both processing mode and software.
Nevertheless, the relatively weak consistency between some solutions using the same broad strategy, such as USNO and AUS, indicates that the distinction between global and independent modes is not sufficient by itself.
Detailed processing choices, including the parameterizations, datum constraints, and software implementation, also contribute to the differences among the coordinate time series.

The NCH-derived precision estimates provide another perspective on these strategy-dependent differences.
For each pair of solutions, we calculated the absolute difference between their NCH-derived precision estimates for the same sources and adopted the median over all sources as a summary statistic, as listed in Table~\ref{tab:precision compare across solutions}.
Similar to the WRMS comparison, this quantity is used as a descriptive measure of the similarity in source-wise precision estimates between two solutions.
The smallest differences are found for the USNO--OPAs pair, with median absolute differences of only $14\,\mathrm{\mu as}$ in right ascension and $22\,\mathrm{\mu as}$ in declination.
Small differences are also found for the USNO--VIE and OPAs--VIE pairs, with values of $29$ and $37\,\mathrm{\mu as}$ in right ascension and declination for the USNO--VIE pair, and $26$ and $37\,\mathrm{\mu as}$ for the OPAs--VIE pair.
This again indicates a high similarity among the USNO, OPAs, and VIE solutions.
The OSO--UAVAC pair also shows small differences, $35\,\mathrm{\mu as}$ in right ascension and $38\,\mathrm{\mu as}$ in declination, suggesting similar source-wise precision levels between these two independent session-wise solutions.
By contrast, the OPAr solution shows much larger differences with most other solutions.
For example, the USNO--OPAr and OPAs--OPAr pairs reach $386$ and $518\,\mathrm{\mu as}$ in right ascension and declination for the former, and $388$ and $523\,\mathrm{\mu as}$ for the latter, respectively.
The comparison between OPAs and OPAr is particularly informative, because both solutions were produced with Calc/Solve but adopted different processing modes.
The much larger differences between OPAs and OPAr suggest that the processing mode has a measurable impact on the NCH-derived precision estimates, even when the same software package is used.

These results show that the similarity of coordinate time-series solutions is not determined solely by the broad classification into global and independent modes.
The close agreement among USNO, OPAs, and VIE, together with the similarity between OSO and UAVAC, indicates that detailed processing choices also play an important role in shaping the scatter and precision characteristics of the coordinate time series.
Therefore, inter-solution comparisons provide an empirical way to assess the realistic precision of source coordinate time series.

\citet{2026A&A...709..A98} provide a method for reducing the effect of random noise by adopting covariance-weighted mean of source positions from multiple catalogs.
For the purpose of deriving coordinate time-series solutions with higher precision, the application of similar approaches is worthy of further investigation.

\subsection{Declination-dependent systematics and precision floor in coordinate time series}

The NCH-derived precision estimates show systematic behavior common to the coordinate time-series solutions from different IVS analysis centers.
As shown in Fig.~\ref{figA:accuracy scatter declination}, most solutions exhibit a clear dependence on source declination.
The precision estimates are generally larger in the southern sky, with a broad maximum around $\delta \simeq -40^{\circ}$, and decrease toward northern declinations.
This pattern is more pronounced in declination than in right ascension.

A similar declination-dependent behavior has been reported for VLBI source catalogs \citep[e.g.,][]{2023A&A...679A..53K}.
The formal errors of ICRF3 source positions also reach larger values around southern declinations and decrease toward the northern sky \citep[e.g.,][]{2020A&A...634A..28L}.
The same general behavior is also consistent with the analysis of source coordinate time series by \citet{2021A&A...648A.125G} and with the results shown by \citet{2018A&A...620A.160L} and \citet{2024ApJS..274...28C}.
The agreement between our NCH-derived time-series precision estimates and previous VLBI catalog analyses suggests that this declination dependence is mainly caused by the observing geometry of the VLBI network rather than by a specific analysis center.

This behavior is expected from the current distribution of VLBI stations.
Most VLBI stations are located in the Northern Hemisphere, and many baselines are predominantly oriented in the east--west direction.
As a result, the north--south component of source positions is less strongly constrained, leading to larger uncertainties in declination.
For southern sources, especially those near and south of $\delta \simeq -40^{\circ}$, the situation is further degraded because they are observed at lower elevations by many northern stations.
The longer atmospheric path length and stronger sensitivity to tropospheric gradients make their coordinate estimates less stable \citep[e.g.,][]{1997GeoRL..24..453M, 2017A&A...606A.143M, 2024ApJS..274...28C}.
Improved observing coverage in the Southern Hemisphere would therefore be important for reducing these declination-dependent systematic effects in future CRF realizations \citep[e.g.,][]{2022Univ....8..374D,2023PASA...40...41W}.

The dependence of the NCH-derived precision on the number of sessions provides another useful diagnostic.
As shown in Fig.~\ref{figA:accuracy scatter sessions}, the typical precision estimates do not decrease systematically as the number of sessions increases.
This behavior suggests the presence of a precision floor in the coordinate time series.
Except for a few poorly constrained NCH estimates, the lower envelope of the NCH-derived precision is about $20\,\mathrm{\mu as}$.
This level is comparable to the noise floor of $30\,\mathrm{\mu as}$ adopted in ICRF3 \citep{2020A&A...644A.159C}.
This agreement suggests that the NCH-derived precision estimates reach a similar lower limit to that found in source positions of VLBI catalogs for well-observed sources.

\subsection{Evaluation of realistic errors in coordinate time series}
\label{sec:Precision and formal error}

\begin{table*}[htbp]
	\centering
    \caption{Comparison between formal errors and NCH-derived precision estimates for 481 sources in right ascension and for 474 sources in declination.}
	\begin{tabular}{lcccccc}
		\hline \hline
		\multirow{2}{*}{Solution} & \multicolumn{3}{c}{RA}                     & \multicolumn{3}{c}{Decl.}                  \\ \cline{2-7}
		                           & below $16\%$ & $(16\%, 84\%)$ & above $84\%$ & below $16\%$ & $(16\%, 84\%)$ & above $84\%$ \\
		(1)                        & (2)           & (3)             & (4)          & (5)           & (6)             & (7)          \\ \hline
		USNO                       & $25.6\%$      & $45.7\%$        & $28.7\%$     & $28.5\%$      & $45.4\%$        & $26.2\%$     \\
		OPAs                       & $28.5\%$      & $49.5\%$        & $22.0\%$     & $33.1\%$      & $45.4\%$        & $21.5\%$     \\
		OPAr                       & $82.5\%$      & $13.1\%$        & $4.4\%$      & $81.6\%$      & $13.3\%$        & $5.1\%$      \\
		AUS                        & $81.1\%$      & $17.7\%$        & $1.2\%$      & $82.1\%$      & $16.0\%$        & $1.9\%$      \\
		OSO                        & $30.4\%$      & $47.4\%$        & $22.2\%$     & $36.9\%$      & $38.4\%$        & $24.7\%$     \\
		VIE                        & $24.7\%$      & $48.6\%$        & $26.6\%$     & $32.1\%$      & $44.7\%$        & $23.2\%$     \\
		UAVAC                      & $42.4\%$      & $44.7\%$        & $12.9\%$     & $50.0\%$      & $38.6\%$        & $11.4\%$      \\
		IAA                        & $9.1\%$       & $23.3\%$        & $67.6\%$     & $12.7\%$       & $23.8\%$        & $63.5\%$     \\ \hline
	\end{tabular}
    \tablefoot{
    Column (1) lists the solution identifier.
    Columns (2) and (5) give the percentages of sources whose median formal errors are smaller than the 16th percentile of the bootstrap distribution of the NCH-derived precision estimates.
    Columns (3) and (6) give the percentages of sources whose median formal errors lie between the 16th and 84th percentiles.
    Columns (4) and (7) give the percentages of sources whose median formal errors are larger than the 84th percentile.}
	\label{tab:compare formal and precision}
\end{table*}

In the construction of celestial reference frames, formal errors of source positions are often rescaled by applying a scaling factor and adding a noise floor, because formal uncertainties derived from least-squares adjustment may decrease approximately as $1/\sqrt{N}$ and therefore underestimate the realistic errors of source positions \citep[e.g.,][]{2014A&A...570A.108L, 2015AJ....150...58F, 2020A&A...644A.159C}.
A similar question arises for coordinate time-series solutions: whether the formal errors of single-session source positions realistically represent the external precision of the time series.

We used the NCH-derived precision estimates as empirical external references to assess the realism of the formal errors.
For each source and each solution, the median formal error was compared with the bootstrap distribution of the corresponding NCH-derived precision estimate.
If the formal error is smaller than the 16th percentile of the NCH-derived precision, it is regarded as underestimated.
If it lies between the 16th and 84th percentiles, it is considered broadly consistent with the NCH-derived precision.
If it is larger than the 84th percentile, it is regarded as conservative.

The results are summarized in Table~\ref{tab:compare formal and precision}.
For the USNO, OPAs, OSO, VIE, and UAVAC solutions, a substantial fraction of sources have formal errors within the 16th--84th percentile interval of the NCH-derived precision estimates.
The fractions are about $45$--$50\%$ in right ascension for USNO, OPAs, OSO, VIE, and UAVAC, and about $45\%$ in declination for USNO, OPAs, and VIE.
This indicates that the formal errors of these solutions are broadly consistent with the external precision estimates for a significant fraction of sources, although underestimation or overestimation is still present for individual sources.

A different behavior is found for the OPAr and AUS solutions.
For these two solutions, more than $80\%$ of sources have formal errors below the 16th percentile of the NCH-derived precision estimates in both coordinates.
The fractions are $82.5\%$ and $81.1\%$ in right ascension for OPAr and AUS, respectively, and $81.6\%$ and $82.1\%$ in declination.
This suggests that the formal errors of OPAr and AUS are systematically underestimated relative to the NCH-derived external precision.
The formal errors of UAVAC in right ascension and declination suffer from similar underestimations, at $42.4\%$ and $50.0\%$, respectively.

The IAA solution shows the opposite behavior.
For IAA, $67.6\%$ of sources in right ascension and $63.5\%$ in declination have formal errors larger than the 84th percentile of the NCH-derived precision estimates.
This indicates that the IAA formal errors are generally conservative compared with the NCH-derived precision.
This behavior is consistent with the relatively large formal errors of the IAA solution reported in Table~\ref{tab:formal error and accuracy}.

These results show that the realism of formal errors differs substantially among the coordinate time-series solutions.
The formal errors of some solutions are broadly consistent with the NCH-derived precision estimates, whereas those of OPAr, AUS, and UAVAC appear to be underestimated and those of IAA appear to be conservative.
This suggests that formal error scaling factor and possible noise floor corrections may also be needed for single-session coordinate time-series solutions, although the appropriate correction should be investigated separately for each analysis strategy and solution type.

\subsection{Stability and sensitivity of the NCH analysis} \label{sect:NCH-shortcoming}

\begin{table*}[htbp]
	\centering
    \caption{Consistency between the 8CH and 7CH NCH-derived precision estimates for 480 sources in right ascension and for 474 sources in declination.}
	\begin{tabular}{lccccccc}
		\hline \hline
		  Statistic    & USNO          & OPAs          & AUS           & OSO           & VIE           & UAVAC         & IAA           \\ \hline
        Median overlap ratio & $31\%$, $30\%$ & $33\%$, $33\%$ & $50\%$, $35\%$ & $62\%$, $67\%$ & $41\%$, $37\%$ & $55\%$, $58\%$ & $42\%$, $39\%$ \\
		Median absolute difference  & 33, 40         & 32, 40         & 22, 29         & 13, 15         & 23, 31         & 15, 18         & 22, 28         \\ \hline
	\end{tabular}
    \tablefoot{All values are given in the order of right ascension and declination.
    The overlap ratio is defined as the length of the overlapping part of the two $68\%$ confidence intervals divided by the total span of their combined interval for each source.
    The median differences are given in micro-arc seconds and rounded to the nearest integer.}
	\label{tab:7CH vs 8CH}
\end{table*}

To evaluate the reliability of the NCH-derived precision estimates, we further examined the stability of the NCH method.
Since eight coordinate time-series solutions were used in the main analysis, this configuration is hereafter referred to as the eight-cornered-hat (8CH) analysis.
After excluding negative variance estimations, physically meaningful NCH estimates were obtained for 481 sources in right ascension and 474 sources in declination.
A source-by-source inspection shows that most of the failed cases have fewer than 100 sessions, suggesting that insufficient observational data are an important cause of the instability.

To test the sensitivity of the NCH results to individual input solutions, we repeated the analysis after excluding the OPAr solution.
This configuration is referred to as the seven-cornered-hat (7CH) analysis.
With the OPAr solution excluded, successful estimates were obtained for 495 sources in right ascension and 496 sources in declination.
The only source that yields negative estimates in the 7CH analysis is 0312+110, which has only ten sessions.
This result indicates that the OPAr solution has a considerable influence on the stability of the original 8CH results.

We then compared the NCH-derived precision estimates obtained from the 8CH and 7CH analyses on a source-by-source basis.
The two analyses have 480 common sources in right ascension and 474 in declination. 
For each common source, we calculated the overlap ratio between the two $68\%$ confidence intervals.
The overlap ratio is defined as the length of the overlapping part of the two intervals divided by the total span of their combined interval.
The median overlap ratio was computed for all common sources.
We also calculated the median absolute difference between the 8CH and 7CH estimates across all common sources.
The results are summarized in Table~\ref{tab:7CH vs 8CH}.

The OSO and UAVAC solutions show the highest consistency between the 8CH and 7CH analyses.
For the OSO solution, the median overlap ratios reach $62\%$ in right ascension and $67\%$ in declination, and the median differences are $13$ and $15\,\mathrm{\mu as}$, respectively.
The corresponding values for UAVAC are $55\%$ and $58\%$ for the overlap ratios and $15$ and $18\,\mathrm{\mu as}$ for the median differences.
The AUS solution also shows a relatively high overlap ratio in right ascension, reaching $50\%$.
By contrast, the USNO, OPAs, VIE, and IAA solutions show smaller overlap ratios, indicating that their NCH-derived precision estimates are more affected by the exclusion of OPAr.
For most sources, the 7CH analysis gives slightly smaller precision estimates than the 8CH analysis.

The effect of excluding OPAr is particularly clear for the two sources with extremely small 8CH precision estimates discussed in Sect.~\ref{subsec:NCH_results}.
For 0537--286, the right ascension precision of OPAs increases to $245\,\mathrm{\mu as}$ in the 7CH analysis, with the 16th and 84th percentiles of $223\,\mathrm{\mu as}$ and $264\,\mathrm{\mu as}$, respectively.
This result is much more stable than the corresponding 8CH estimate.
For source 0823--223, the right ascension precision of IAA is estimated to be $205\,\mathrm{\mu as}$ in the 7CH analysis, with the 16th and 84th percentiles of $-150\,\mathrm{\mu as}$ and $366\,\mathrm{\mu as}$, respectively.
The negative lower bound indicates that this estimate remains poorly constrained, mainly because only 11 sessions are available for this source.

These tests show that the stability of the NCH analysis depends on both the consistency among the input solutions and the number of common sessions.
For sources with a small number of sessions, the NCH estimates can become unstable because the pairwise WRMS values and correlation coefficients are poorly constrained.
For some well-observed sources, large discrepancies between the 8CH and 7CH results are instead associated with large scatter in the pairwise offset time series (e.g., Fig.~\ref{figA:same session offset from USNO for 0923+392}).
Such large scatter can affect the estimation of the correlation coefficients and may lead to noticeable differences between the 8CH and 7CH NCH-derived precision estimates.

\section{Conclusions}
\label{sec:Conclusion}

We collected eight coordinate time-series solutions of extragalactic radio sources from seven IVS analysis centers, including USNO, OPAs, OPAr, AUS, OSO, VIE, UAVAC, and IAA.
These solutions were produced using different processing strategies, software packages, parameterizations, and source constraints.
After data screening and the extraction of common sessions, the final sample contains 496 sources observed in at least ten common sessions.
Based on this sample, we compared the coordinate time-series solutions, evaluated their external consistency, and estimated the positional precision of each source in each solution using the NCH method.

The coordinate time-series solutions show both common features and solution-dependent differences.
For the same source and the same VLBI session, the estimated positions can differ among analysis centers, reflecting the influence of different processing strategies.
Pairwise offsets between solutions largely reduce common-mode variations and reveal the inter-solution differences more clearly.
For most solutions, the typical WRMS values of the coordinate offsets are at the level of a few hundred micro-arc seconds in both right ascension and declination.

The comparison among different solutions shows that processing strategies have a clear impact on the stability and consistency of the coordinate time series.
Solutions produced with similar strategies or constraints show more similar WRMS values, stronger correlations, and closer NCH-derived precision estimates.
In particular, USNO and OPAs show the highest consistency, which is likely related to their common use of the global mode and the Calc/Solve software.
The OSO and UAVAC solutions also show similar behaviors, especially in their WRMS and NCH-derived precision estimates, likely because both are based on independent session-wise analyses.
The VIE solution, although produced in an independent mode, shows closer agreement with USNO and OPAs than with the other independent solutions, likely because the subsets of the ICRF3-defining sources were fixed in its two complementary solutions.

The NCH-derived precision estimates provide an independent assessment of the realistic precision of the coordinate time series.
For most solutions, the median NCH-derived precision is approximately $200$--$300\,\mathrm{\mu as}$ in right ascension and $250$--$400\,\mathrm{\mu as}$ in declination.
The OPAr solution shows the largest NCH-derived precision estimates in both coordinates, indicating poorer external precision compared to the other solutions.
The formal errors are broadly consistent with the precision estimates derived from NCH for several solutions, but clear differences are found for OPAr, AUS, UAVAC, and IAA.
The formal errors of OPAr, AUS, and UAVAC appear to be underestimated relative to the NCH-derived precision, whereas those of IAA appear to be conservative.

The NCH-derived precision estimates also show a systematic dependence on the source declination.
For most solutions, the precision estimates are larger in the southern sky and decrease toward northern declinations, with a broad maximum around $\delta \simeq -40^{\circ}$.
This pattern is consistent with previous analyses of VLBI source catalog uncertainties and is mainly related to the geometry of the VLBI observing network.
The predominance of Northern-Hemisphere stations and the limited observing geometry for southern sources lead to weaker constraints, particularly in declination.
Improved Southern-Hemisphere observing coverage would therefore help reduce these declination-dependent systematic effects.

The stability of the NCH analysis was further examined by comparing the original 8CH analysis with a 7CH analysis in which the OPAr solution was excluded.
The exclusion of OPAr increases the number of sources with successful NCH estimates, indicating that OPAr has a considerable impact on the stability of the NCH results.
For sources with a few common sessions, the NCH estimates can become unstable.
For some well-observed sources, large scatter in the pairwise offset time series can also lead to noticeable differences between the two NCH analyses.

Overall, this study focuses on inter-solution comparisons and provides quantitative metrics to assess the external consistency and realistic precision of source coordinate time series.
Such precision estimates are important for studies based on VLBI source time series, including investigations of source-position stability, astrometric variability related to source structure, and the long-term maintenance of the CRF.

\begin{acknowledgements}
We sincerely thank the referee for their careful review and constructive comments, and the editor for their professional guidance.
N.L. is supported by the Astrometric Reference Frame project under grant No.~JZZX-0102.
N.L. and J.C.L are also supported by the National Natural Science Foundation of China (NSFC) under grant Nos.~12573070 and 12373074.
This research used data provided by the United States Naval Observatory IVS Analysis Center, the Paris Observatory Geodetic VLBI Center, Geoscience Australia, the Onsala Space Observatory, the Vienna IVS Analysis Center and the Universidad de Alicante VLBI Analysis Center.
We express our special thanks to Renata Urunova at the Institute of Applied Astronomy of the Russian Academy of Sciences for kindly providing the data used in this work.
We sincerely thank Angelina Osetrova for the fruitful discussions and valuable suggestions that have substantially improved this work.
We made extensive use of Astropy \citep[][\url{http://www.astropy.org}]{2018AJ....156..123A} and the Python 2D plotting library Matplotlib \citep{2007CSE.....9...90H}.
We also appreciate the LaTeX and Git services provided by the e-Science Center of the Collaborative Innovation Center of Advanced Microstructures, Nanjing University (\url{https://sci.nju.edu.cn}).
The Jupyter Notebook used to generate the figures for all 496 sources in the final sample of this paper is available at \url{https://git.nju.edu.cn/astrometry/vlbi_src_ts_precision} and is preserved on Zenodo at \url{https://doi.org/10.5281/zenodo.22674551}.

\end{acknowledgements}

\bibliographystyle{aa} 
\bibliography{reference}

\begin{appendix}
\nolinenumbers
\onecolumn

\section{Temporal variation of coordinate time-series solutions for 0923+392}
\label{appendix:coordinate_vs_t}

\begin{figure}[hbtp]
    \centering
    \includegraphics[width=0.49\textwidth]{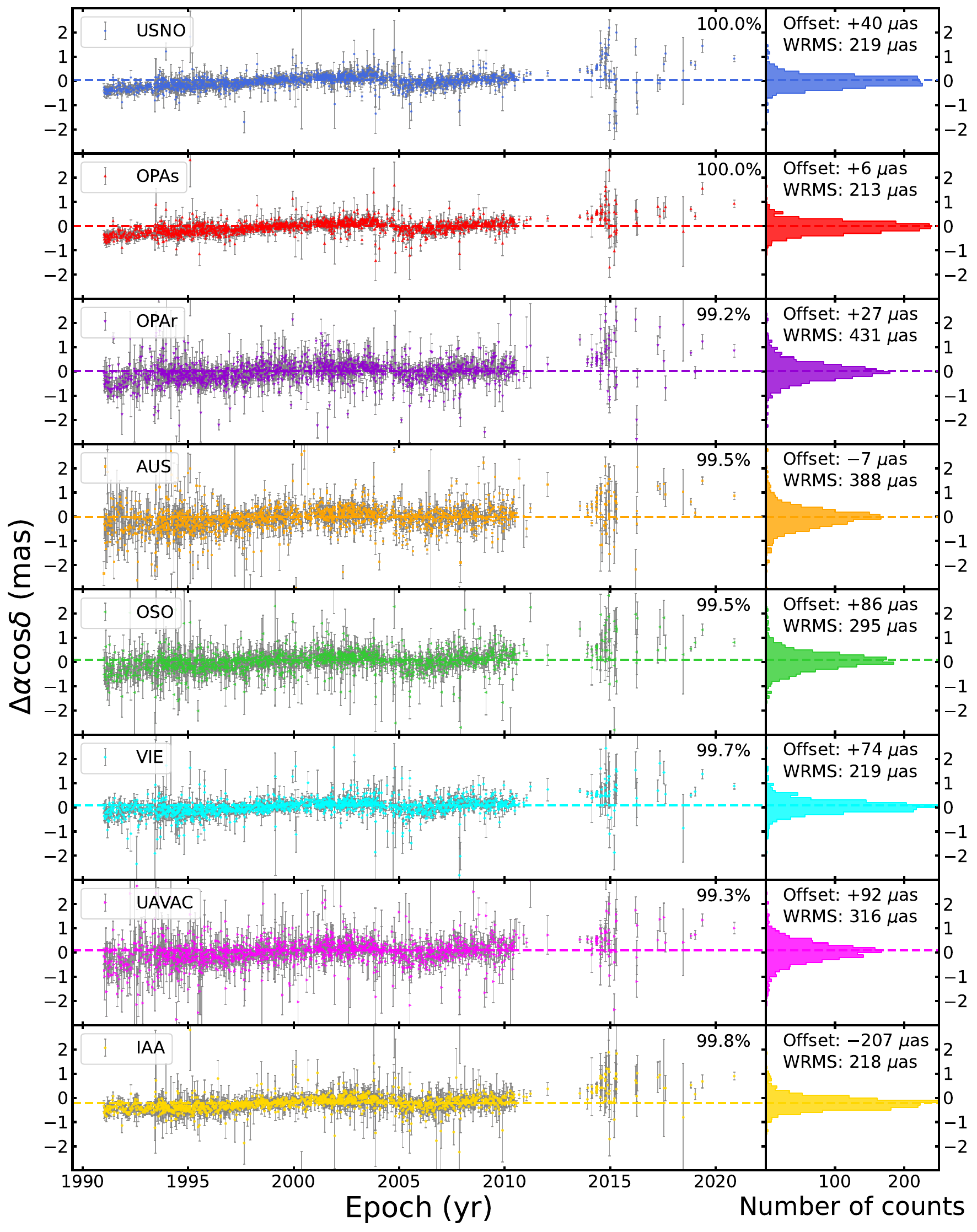}
    \includegraphics[width=0.49\textwidth]{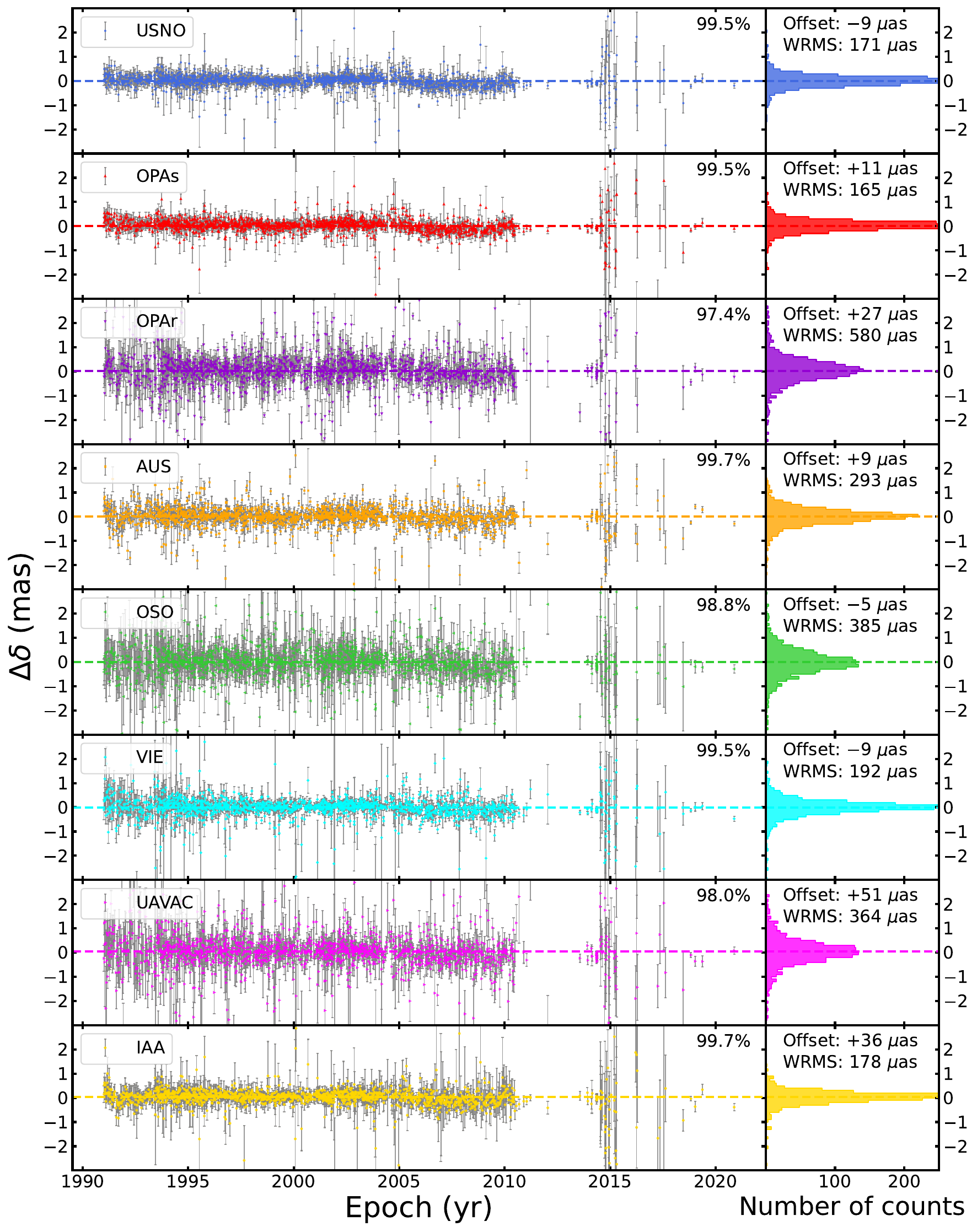}
    \caption{Eight source coordinate time-series solutions and their formal errors for 0923+392, shown separately for right ascension (left) and declination (right). 
    In each panel, each row corresponds to one time-series solution, with the solution name marked in the top left corner of each scatter plot.
    The reference value is the position given in the ICRF3 catalog. 
    The number in the top right corner of each scatter plot indicates the percentage of data points lying between $-3$ and $3$ mas.
    The distributions of the offsets within this range are shown by histograms with 60 bins.
    The offset given in the top right corner of each histogram represents the difference between the ML position, indicated by the dashed line, and the ICRF3 position. The WRMS is computed from each offset series according to Eq.~(\ref{Equ:WRMS}).
    The offset and WRMS displayed are rounded to nearest integer.}
    \label{figA:Time series solutions for 0923+392}
\end{figure}

\begin{figure}[htbp]
  \centering
  \includegraphics[width=\textwidth]{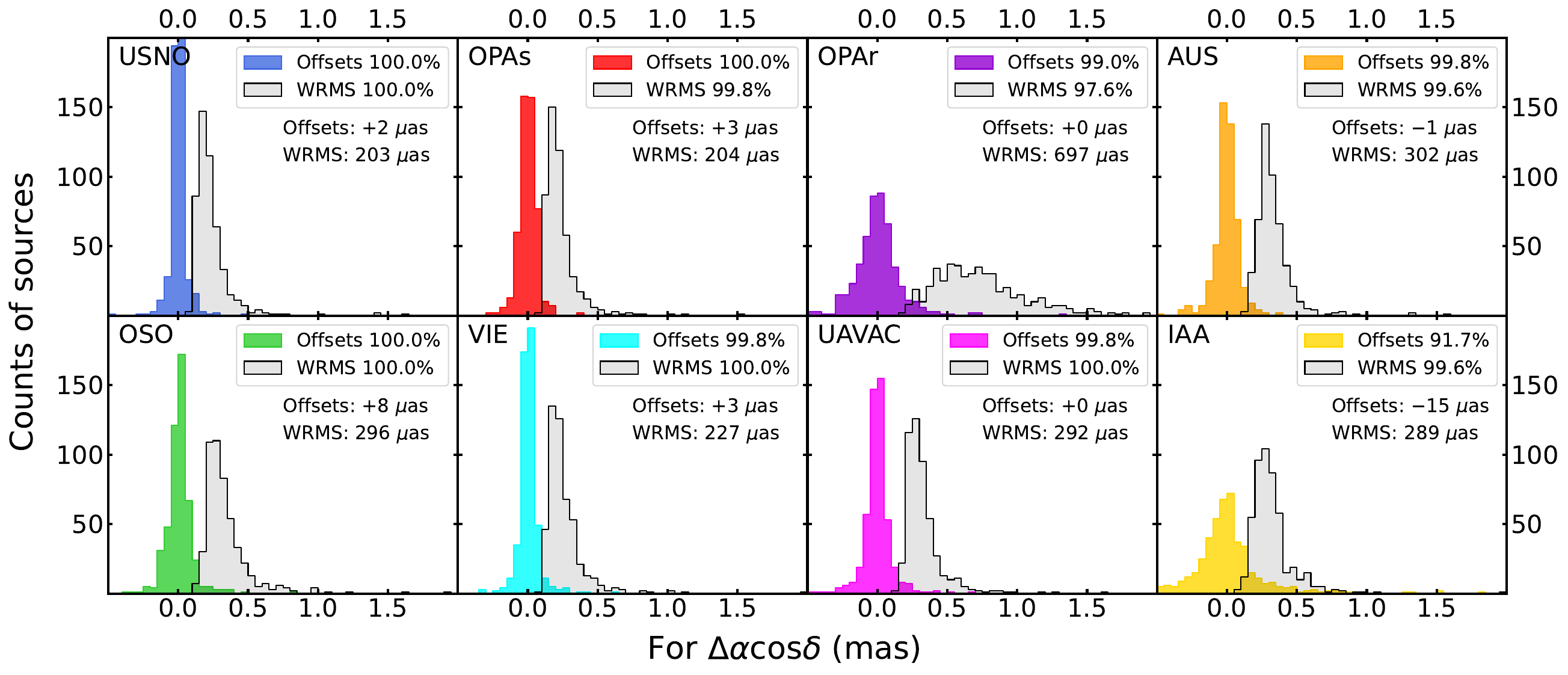}
  \includegraphics[width=\textwidth]{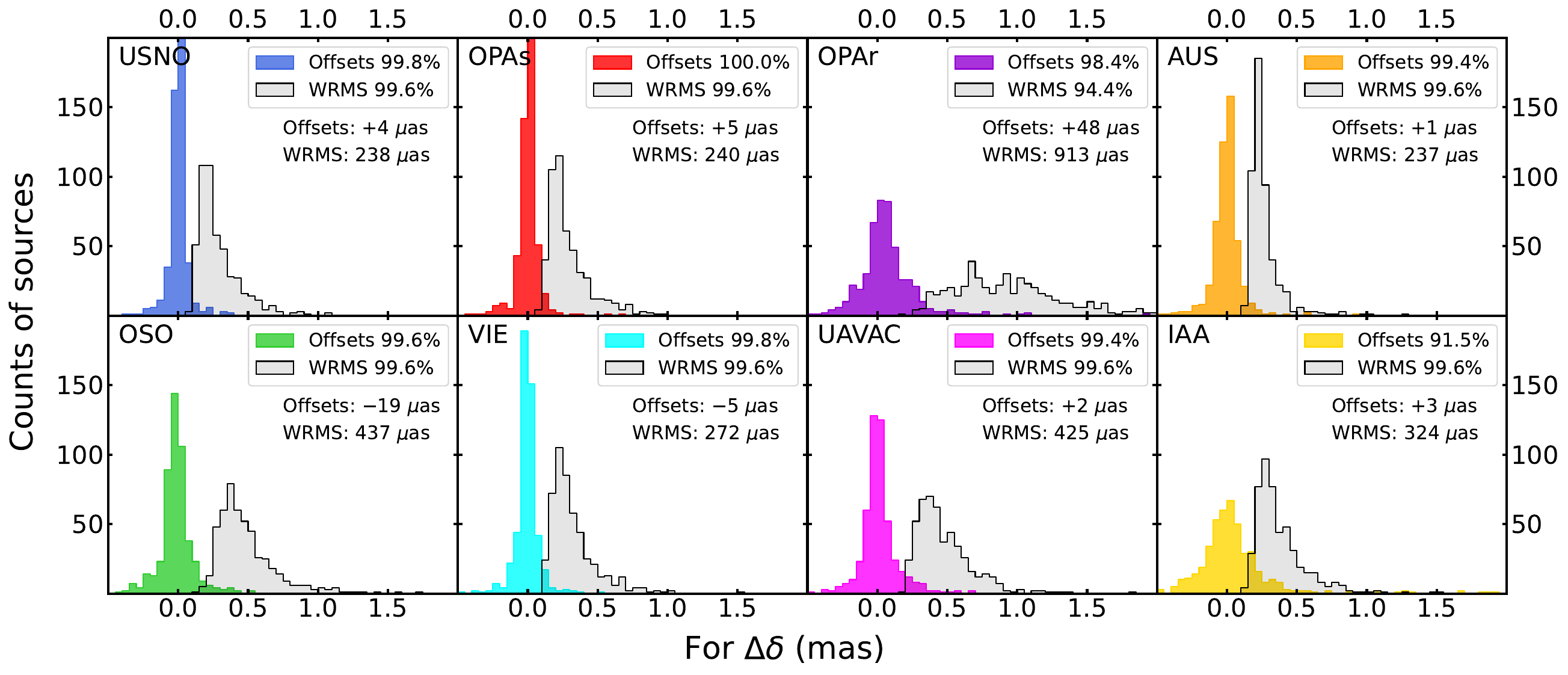}
  \caption{Statistics of offsets and WRMS calculated in Fig.~\ref{figA:Time series solutions for 0923+392} for 496 sources in the final sample from different solutions, for right ascension (top) and declination (bottom).
  The range of offsets or WRMS for plots are limited to the range between $-0.5$ and $2$ mas, and the percentages in each label indicate the number of data points lying in this range.
  The distributions of the number of counts within this range are indicated with histograms with 50 bins.
  The offset and WRMS given in the top right corner of each panel represent the medians of their respective distributions.}
  \label{figA:statistics of offset and WRMS}
\end{figure}

As shown in Fig.~\ref{figA:Time series solutions for 0923+392}, the right ascension and declination time series for source 0923+392 exhibit similar nonlinear variation patterns across all solutions.
This occurs despite the slight systematic differences in the ML positions and the position estimates for the corresponding sessions. 
Notably, the OPAr, AUS, OSO, and UAVAC solutions display larger scatter and higher WRMS values, whereas solutions from the remaining centers demonstrate more compact behavior.

A pronounced positional shift in both right ascension and declination is visible around 2015 across all solutions. 
Furthermore, the right ascension time series reveals quasiperiodic variations, featuring peaks around 2004 and 2010 alongside troughs near 1996 and 2005. 
While these variations may be linked to the intrinsic physical properties of the source, the fluctuations in declination are noticeably less pronounced. 
The declination coordinate displays potential peaks around 1993 and 2005 and troughs near 1999 and 2007, superimposed on a weak, long-term decline.
The weak linear positional variations in right ascension before 1995 and a small peak in declination near 1993 agree well with the results from \citet{1997AJ....114.2284F}.
The underlying physical mechanisms could be related to the twisted, relativistic jet model and the superluminal motion of component with respect to the stationary components from west to east \citep[e.g.,][]{1991ApJ...371..491M, 1993ApJ...402..160A, 1995AJ....110.2586G, 1999NewAR..43..731A, 2000A&A...361..529A}.

Figure~\ref{figA:statistics of offset and WRMS} displays the distribution of offsets and WRMS values for all sources.
For the offset distribution, all solutions exhibit near-Gaussian distributions, with all peaks centered close to zero, for both right ascension and declination.
In particular, the right ascension peaks of USNO and VIE are prominent, while the declination distributions of USNO and OPAs also show distinctly the highest peaks.
The offset distributions in both right ascension and declination of the IAA solution show the lowest peaks.
For the WRMS distributions, all solutions show non-Gaussian distributions.
The OPAr solution exhibits a much broader WRMS distribution than the other solutions in both right ascension and declination.
In right ascension, the peaks of WRMS distributions for USNO, OPAs and VIE are around 0.2\,mas, while for AUS, OSO, UAVAC and IAA are approximately at 0.3\,mas.
In declination, the peak positions of all solutions apart from AUS and IAA show a slight increase.
For USNO, OPAs and VIE, the peaks are around 0.25\,mas, while those of OSO and UAVAC are approximately 0.4\,mas.
The peak of AUS shifts leftward, whereas the peak position of IAA remains nearly unchanged.
\clearpage
\section{Correlation among coordinate time-series solutions for 0923+392} \label{secA:Consistency}

\begin{figure}[hbtp]
    \centering
    \includegraphics[width=0.49\textwidth]{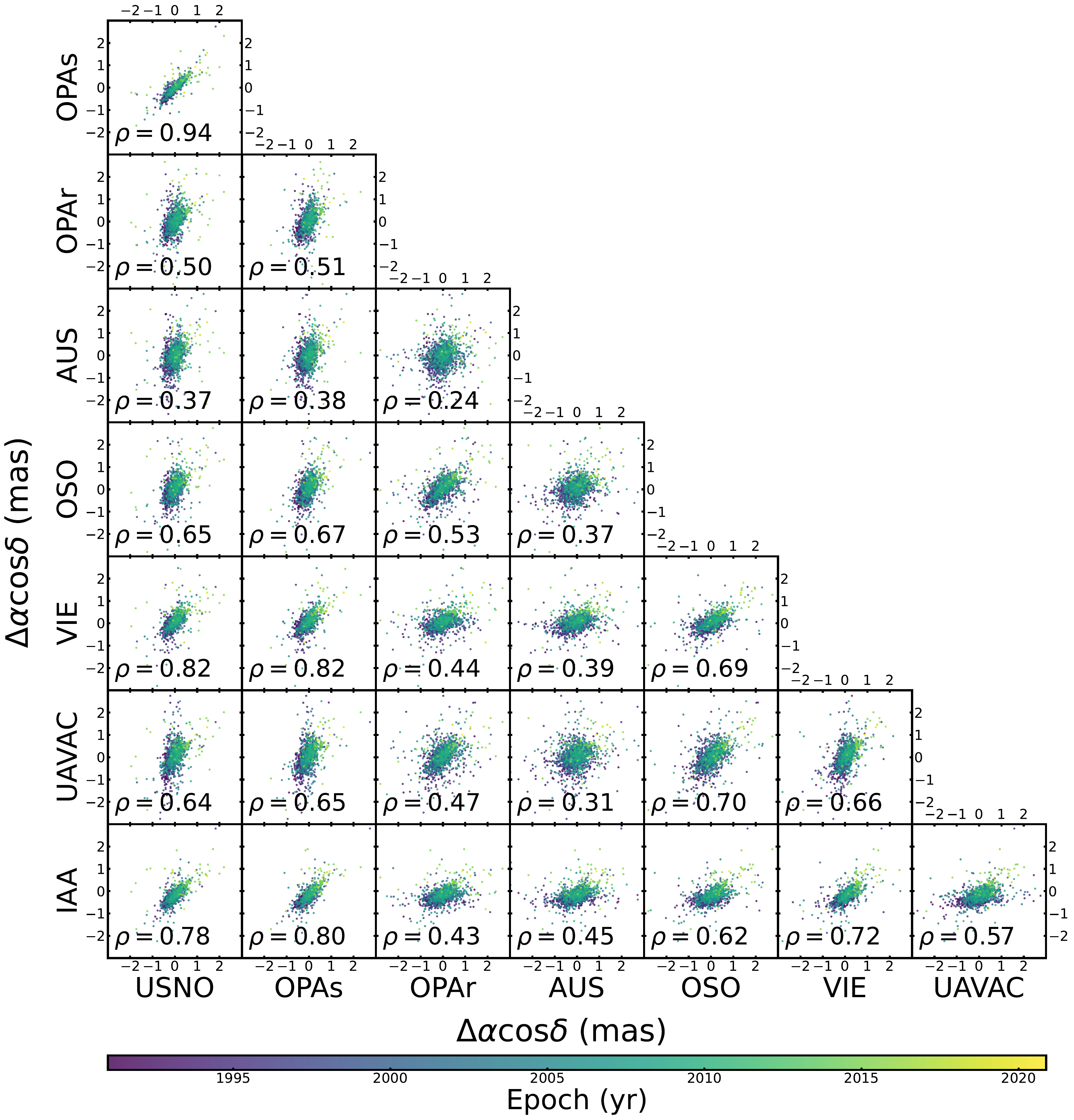}
    \includegraphics[width=0.49\textwidth]{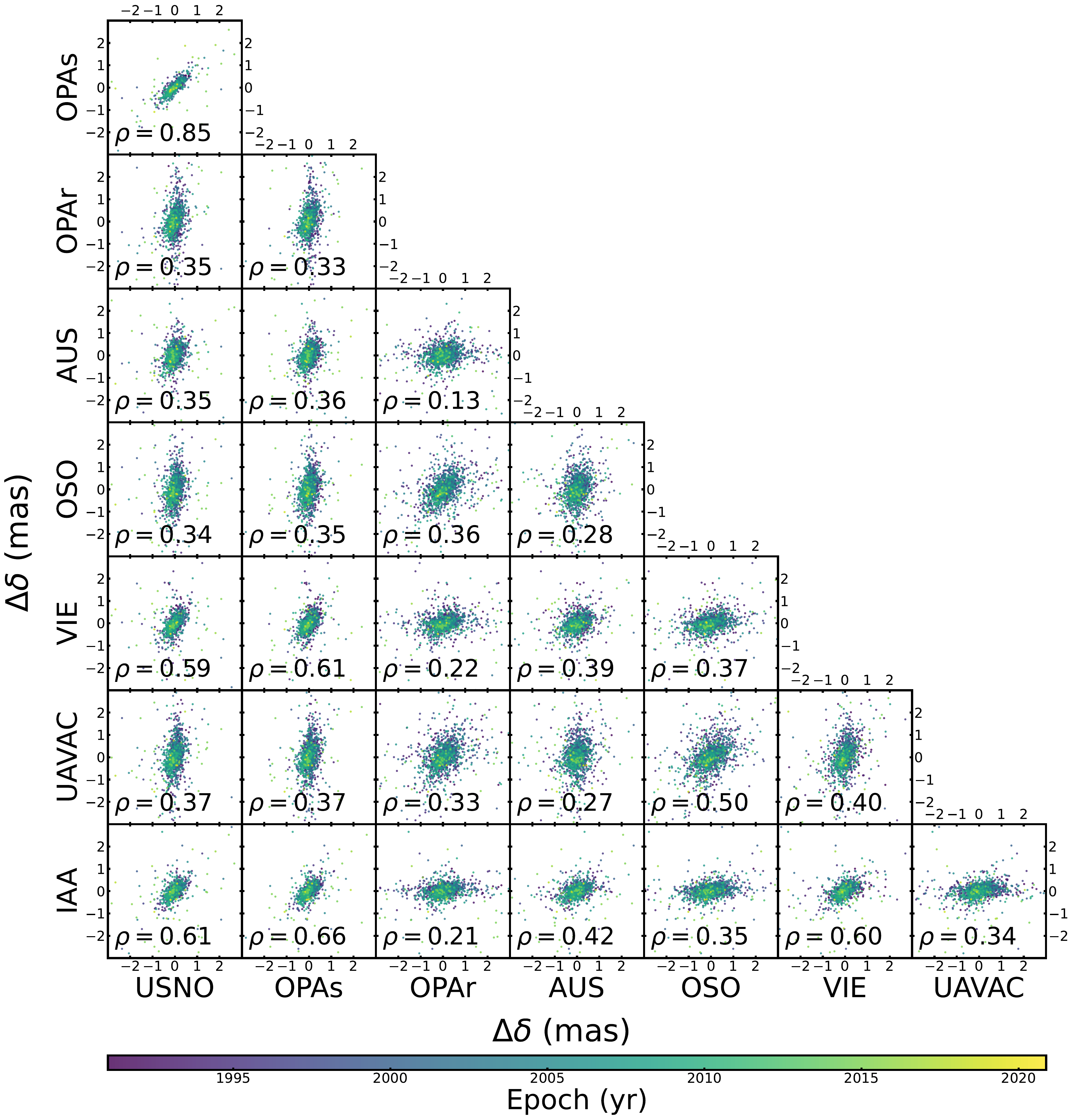}
    \caption{Correlations between each pair of coordinate time-series solutions for 0923+392, shown separately for right ascension (left) and declination (right). 
    In each subplot, the positions from the same sessions in the two time-series solutions are shown as colored dots, with the color indicating the session epoch.
    The reference point corresponds to the position given in the ICRF3 catalog; the axis labels are therefore $\Delta \alpha \cos \delta$ and $\Delta \delta$.
    The correlation coefficient for each pair is computed using Eq.~(\ref{equ:correlation}) and is given in the bottom left corner of each subplot.}
    \label{figA:correlation 0923+392}
\end{figure}

Figure~\ref{figA:correlation 0923+392} depicts the Pearson correlation coefficients for each pair of solutions for 0923+392.
For both right ascension and declination, all pairs show positive correlations.
In right ascension, the weakest correlation is found for the OPAr--AUS pair, with $\rho=0.24$, whereas the strongest correlation, $\rho=0.94$, is found for the OPAs--USNO pair.
Most of the other pairs have Pearson correlation coefficients of about $0.5$.
For the declination coordinate, the correlation coefficient is generally smaller than that of the corresponding right ascension pair, except for the AUS--VIE pair, for which the correlation is 0.39 in both right ascension and declination.
The weakest correlation is again found for the OPAr--AUS pair, with $\rho=0.13$, indicating an almost negligible correlation in declination.
The strongest correlation is again found for the OPAs--USNO pair, with $\rho=0.85$.
\clearpage
\section{Comparison of NCH-derived precision with formal errors}
\label{secA:Comparison between precision and formal errors}

\begin{figure}[hbtp]
    \centering
    \includegraphics[width=0.49\textwidth]{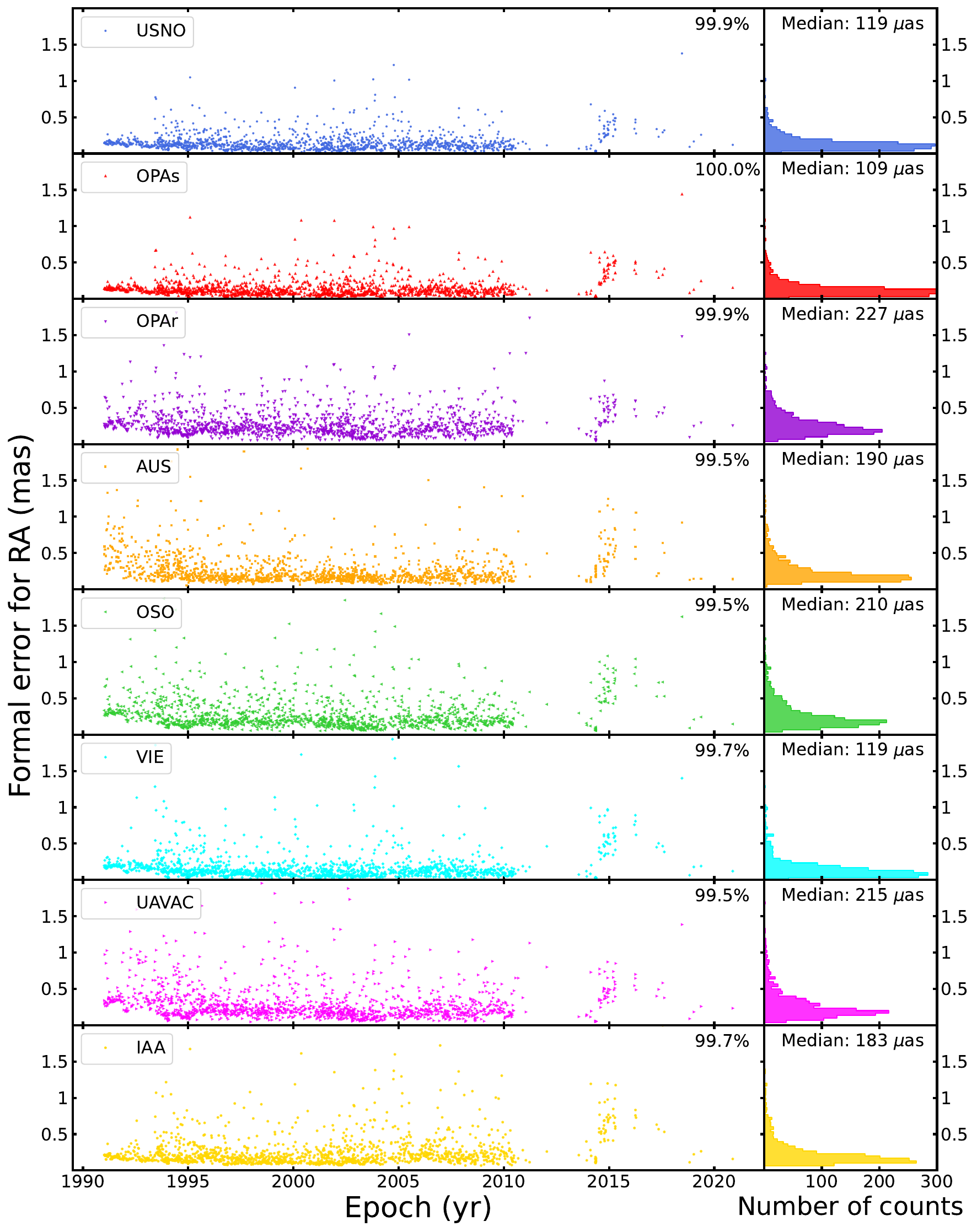}
    \includegraphics[width=0.49\textwidth]{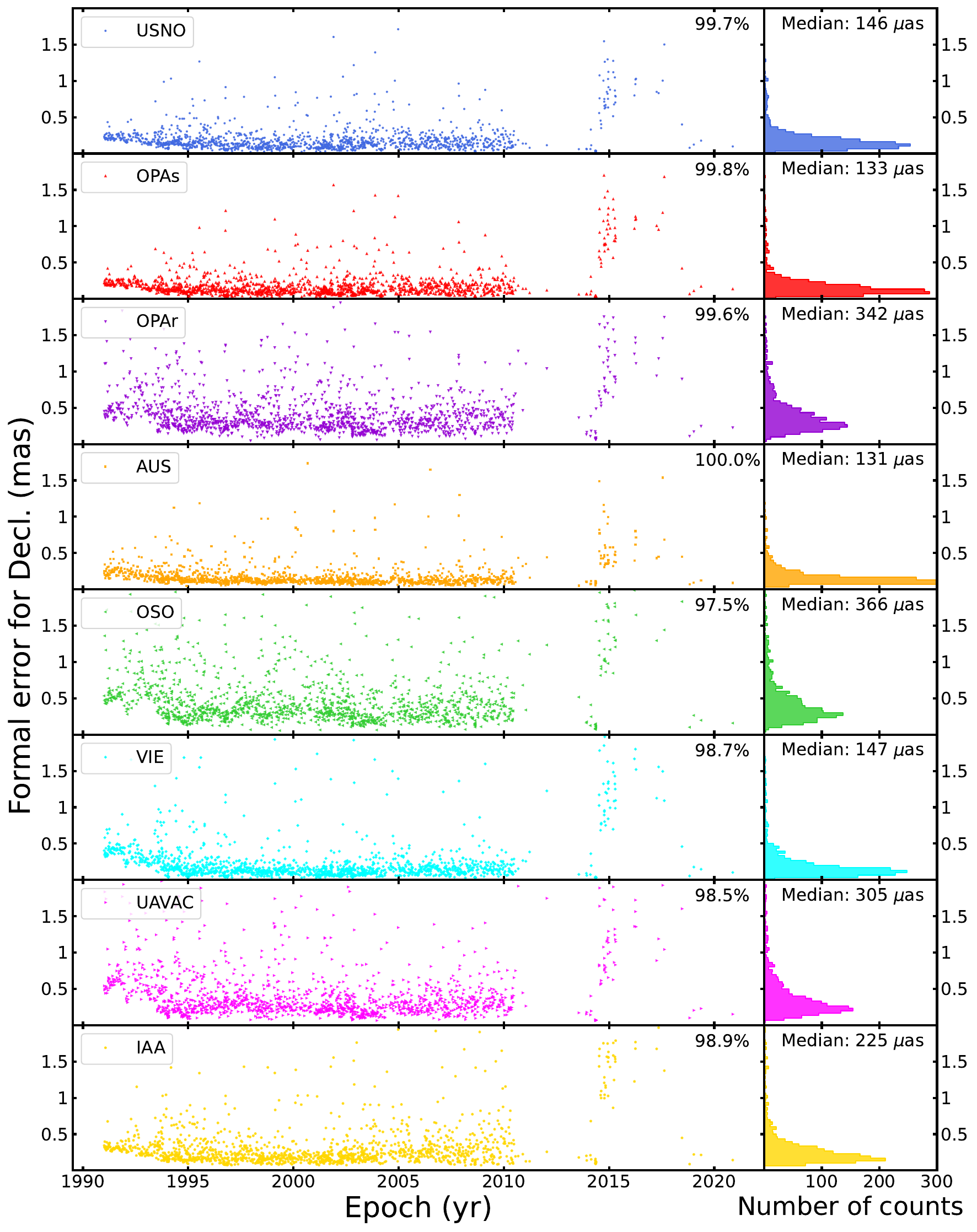}
    \caption{Formal errors of eight time-series solutions for 0923+392, shown separately for right ascension (left) and declination (right).
    The median for each formal error time series is displayed in the top right corner of the corresponding histogram.}
    \label{figA:formal error for 0923+392}
\end{figure}
\begin{figure}[htbp]
  \centering
  \includegraphics[width=\textwidth]{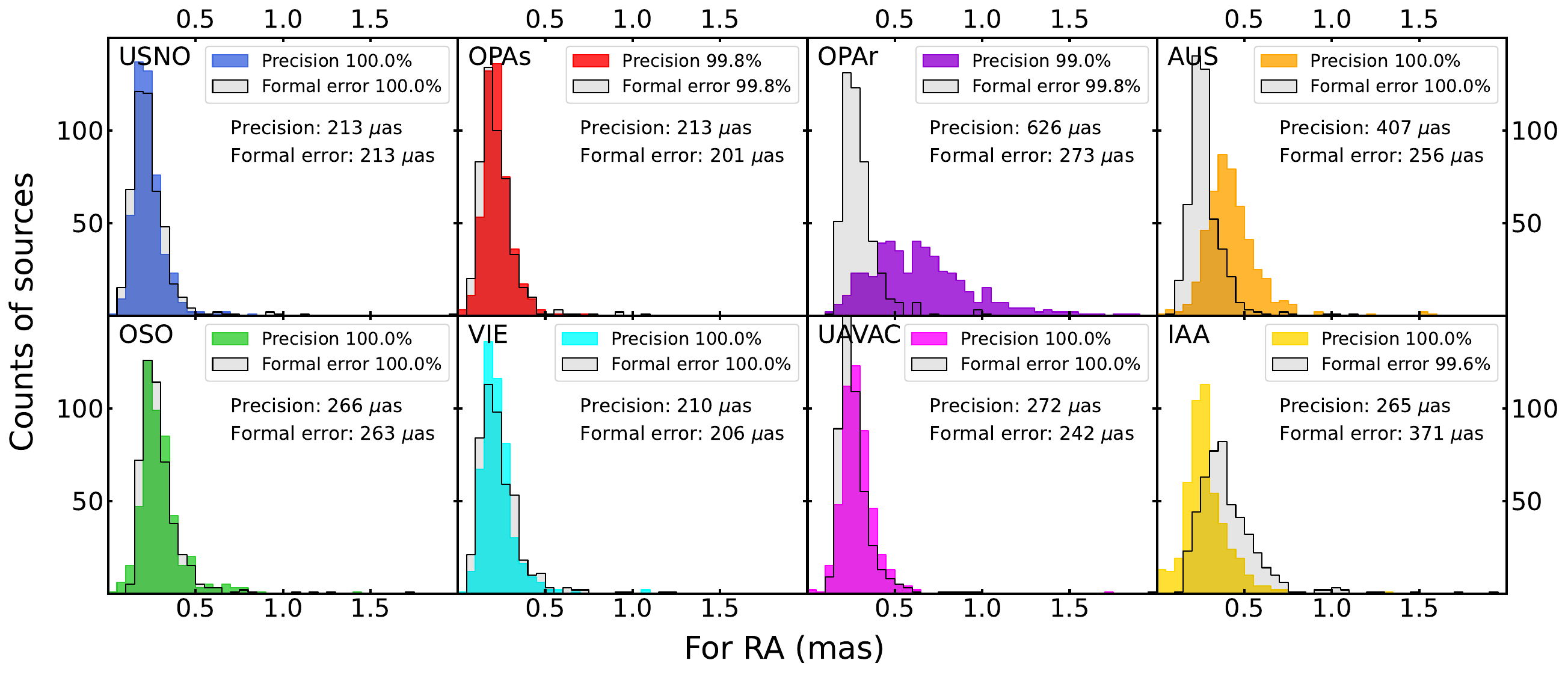}
  \includegraphics[width=\textwidth]{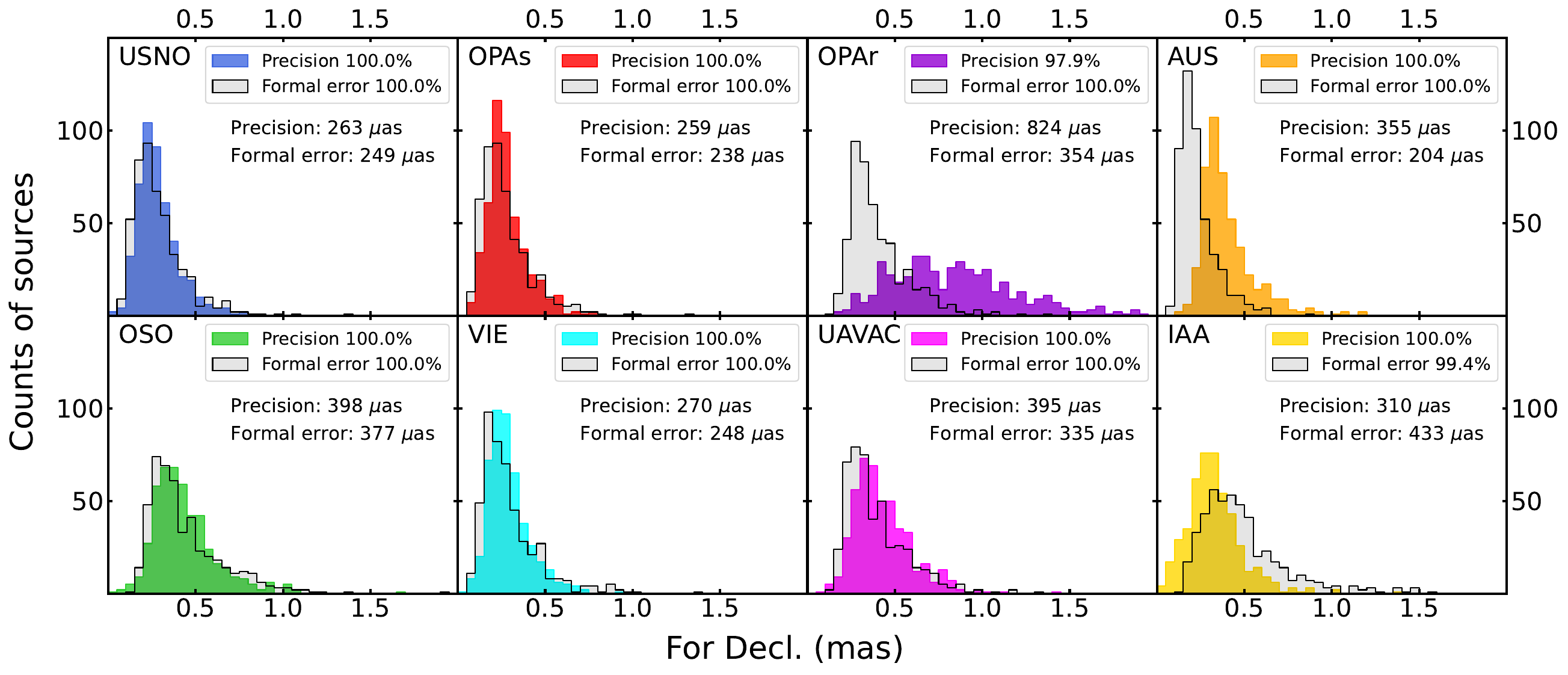}
  \caption{Precision and median of the formal errors for the sources from different time-series solutions (represented by different colors), for 481 sources in right ascension (top) and for 474 sources in declination (bottom).
  Precision or formal errors for plots are limited to the range between 0 and 2 mas.
  The distributions of data points within this range are indicated with histograms with 40 bins, and the number of counts are presented in the label.
  The median of precision and formal error are given in the top right corner of each panel.}
  \label{figA:comparison}
\end{figure}

The formal errors of 0923+392 for individual sessions are presented in Fig.~\ref{figA:formal error for 0923+392}.
In both right ascension and declination, most formal errors lie between 0 and 1\,mas, with a significant decreasing trend before 1994.
The increase in formal errors around 2015 is consistent with the significant positional variations shown in Fig.~\ref{figA:Time series solutions for 0923+392}.
For each solution, the formal error in right ascension is generally smaller than that in declination, except for AUS.
The degree of data point scatter across all solutions is similar, while the formal errors for the three independent mode OPAr, OSO, and UAVAC are noticeably larger than those of the other solutions, especially in declination.

Figure~\ref{figA:comparison} compares the distributions of NCH-derived precision and formal errors. 
For all solutions, both quantities exhibit heavy-tailed, non-Gaussian distributions in right ascension and declination.
For precision, the most dispersed distribution can be observed in OPAr.
In right ascension, the peaks of precision for USNO, OPAs and VIE are located near 0.2\,mas, while for OSO, UAVAC and IAA are around 0.25\,mas, for AUS is approximately at 0.4\,mas.
In declination, similarly, except for AUS, the peak positions exhibit a slight increase for most solutions.
The peaks are at about 0.25\,mas for USNO, OPAs, and VIE, and around 0.4\,mas for OSO and UAVAC.
For AUS, the peak position shows a decreasing behavior and the peak of IAA remains around 0.25\,mas.
Both precision and formal error exhibit lower peaks and larger scatter in declination than that in right ascension for each solution.
\clearpage
\section{Precision of source coordinate time series derived from NCH method}
\label{secA:NCH results}

\begin{figure}[htbp]
  \centering
  \includegraphics[width=\textwidth]{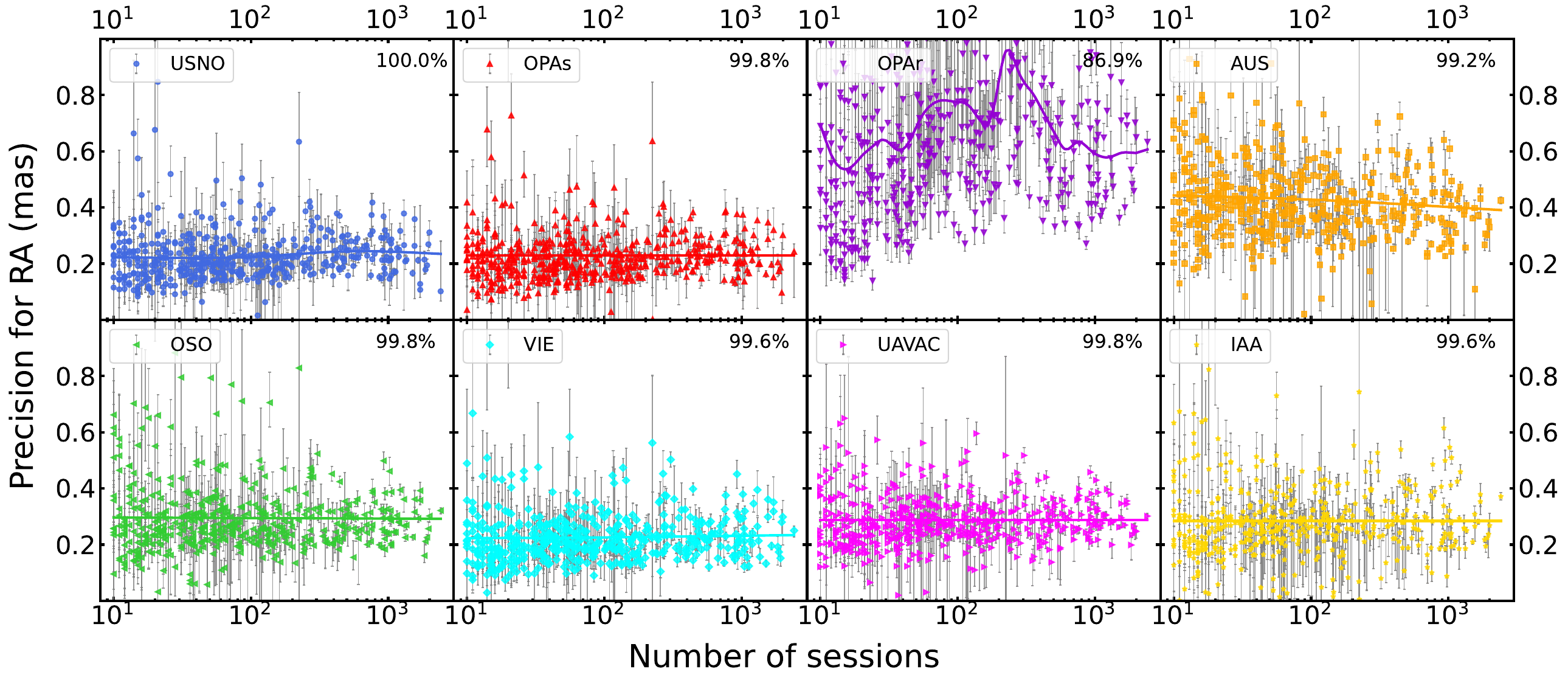}
  \includegraphics[width=\textwidth]{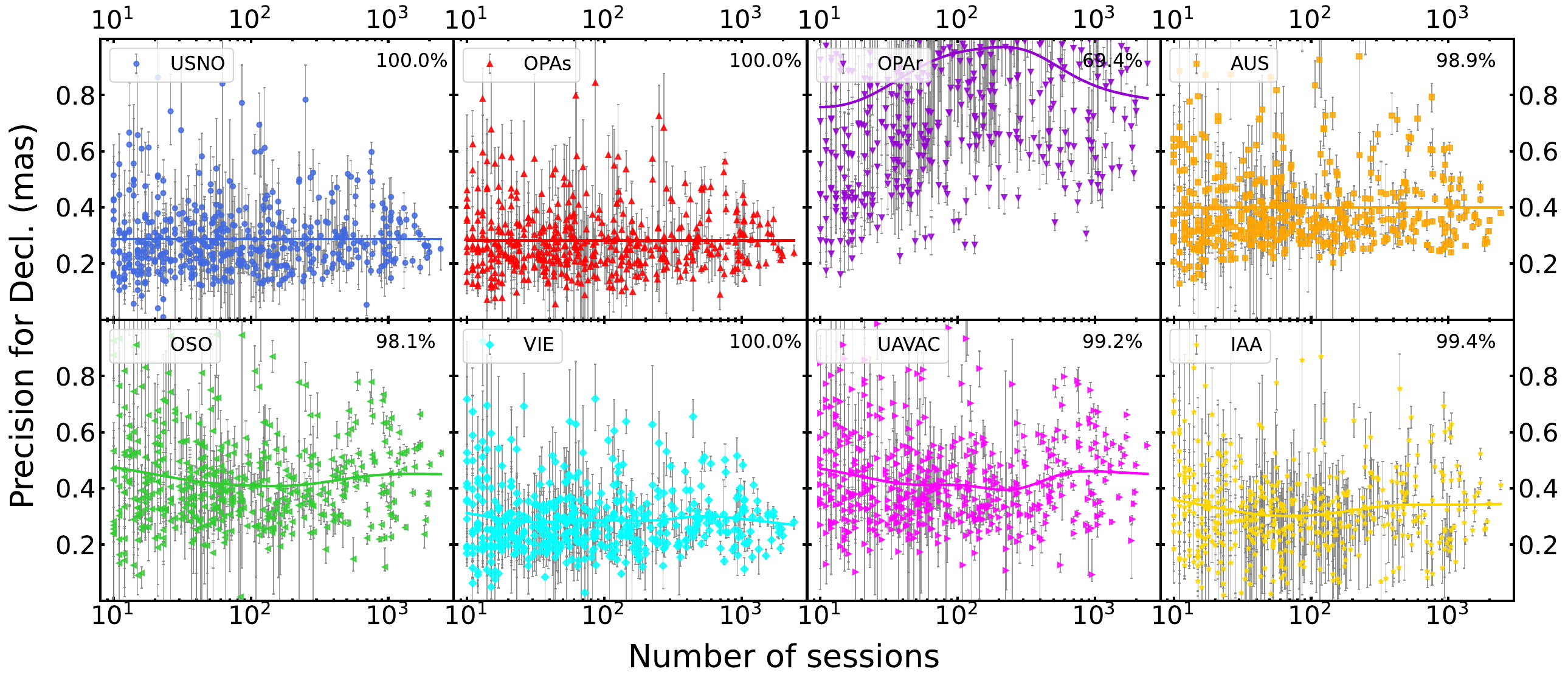}
  \caption{NCH-derived precision for each source from different time-series solutions (color-coded by solution) vs. the number of sessions, shown separately for 481 sources in right ascension (top) and for 474 sources in declination (bottom).
  The error bars represent the 16th and 84th percentiles.
  The precision range in each panel is restricted to 0 to 1\,mas, and the number of data points falling within this interval is shown in the top right corner of each panel.
  The curves show smoothed trends.}
  \label{figA:accuracy scatter sessions}
\end{figure}

\begin{figure}[htbp]
  \centering
  \includegraphics[width=\textwidth]{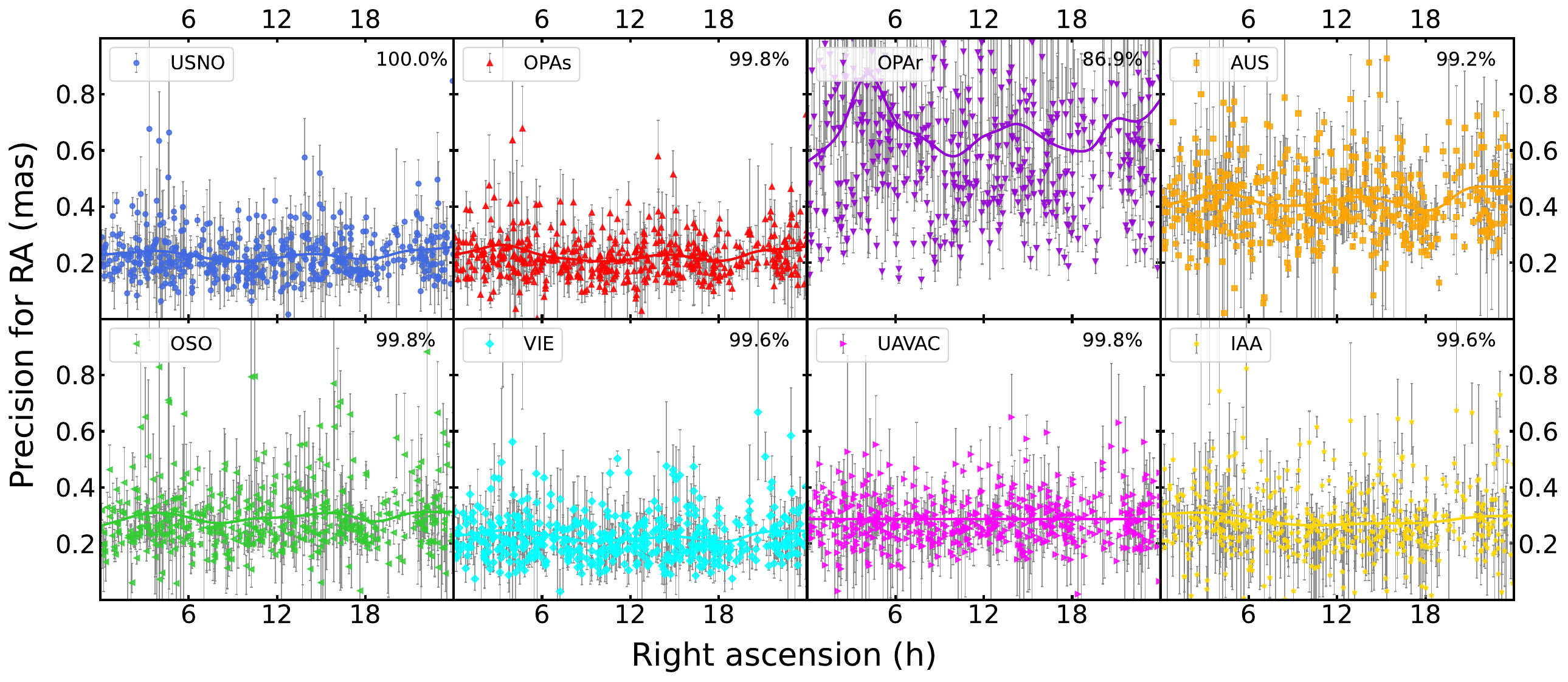}
  \includegraphics[width=\textwidth]{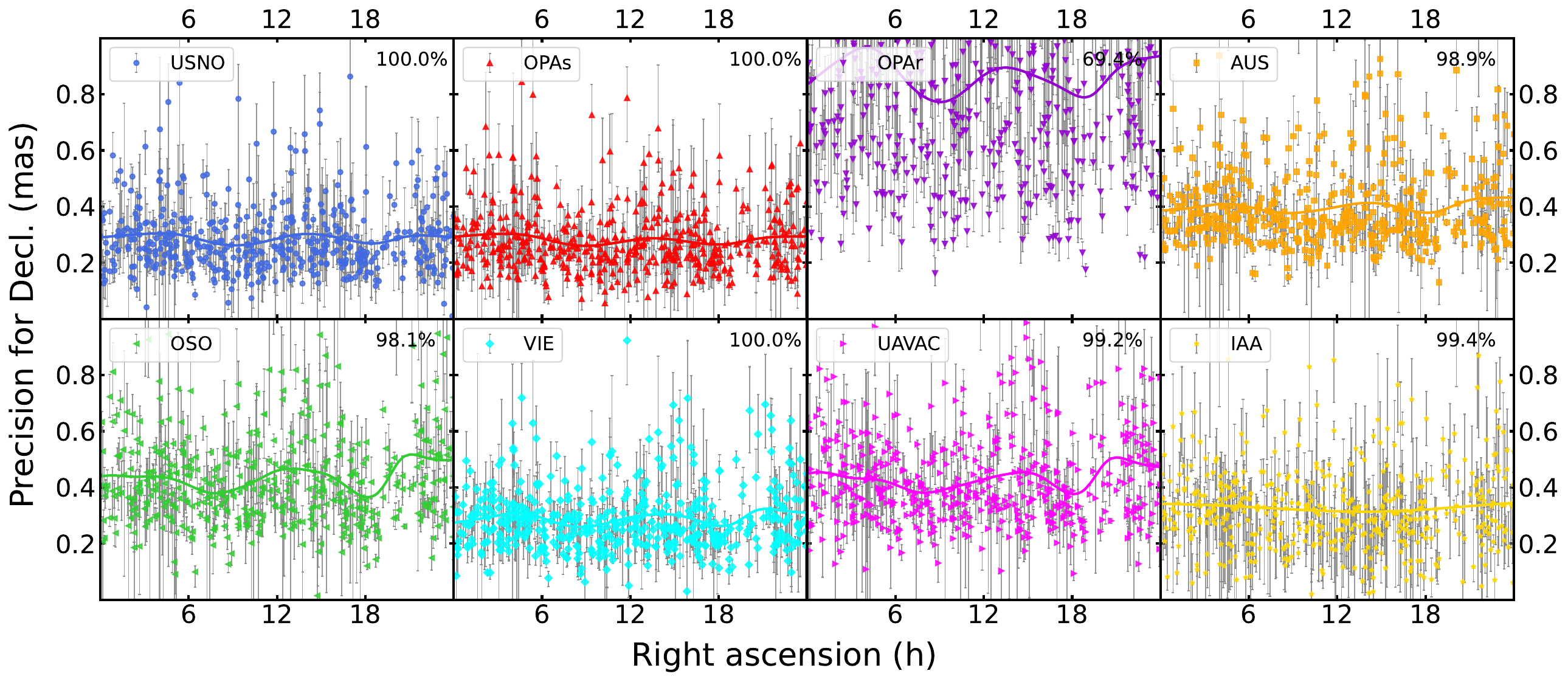}
  \caption{NCH-derived positional precision of individual sources derived from different time-series solutions (color-coded) as a function of right ascension for 481 sources in right ascension (top) and for 474 sources in declination (bottom).
  The error bars correspond to the 16th and 84th percentiles.
  The value displayed in the top right corner of each panel gives the percentage of data points lying within the 0 to 1\,mas precision interval.
  The curves show smoothed trends.}
  \label{figA:accuracy scatter right ascension}
\end{figure}

\begin{figure}[htbp]
  \centering
  \includegraphics[width=\textwidth]{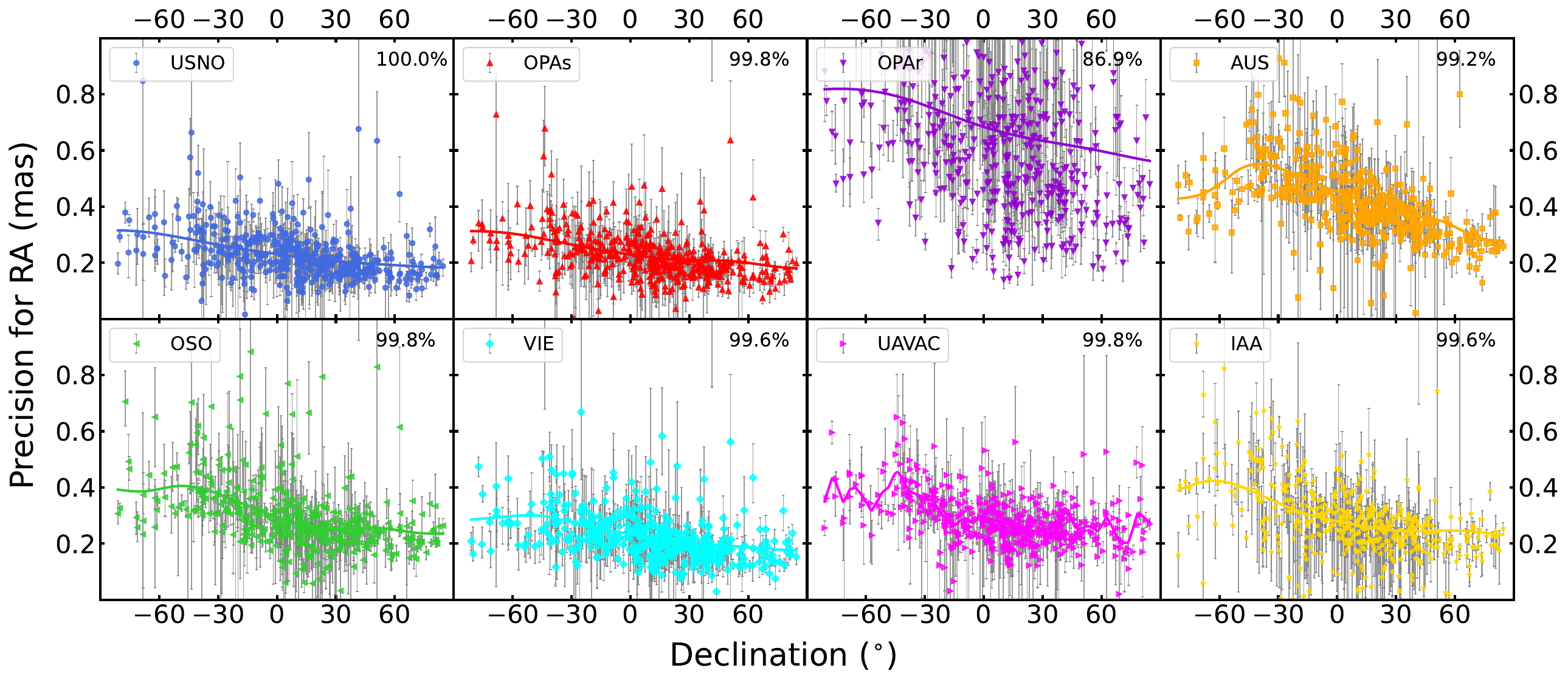}
  \includegraphics[width=\textwidth]{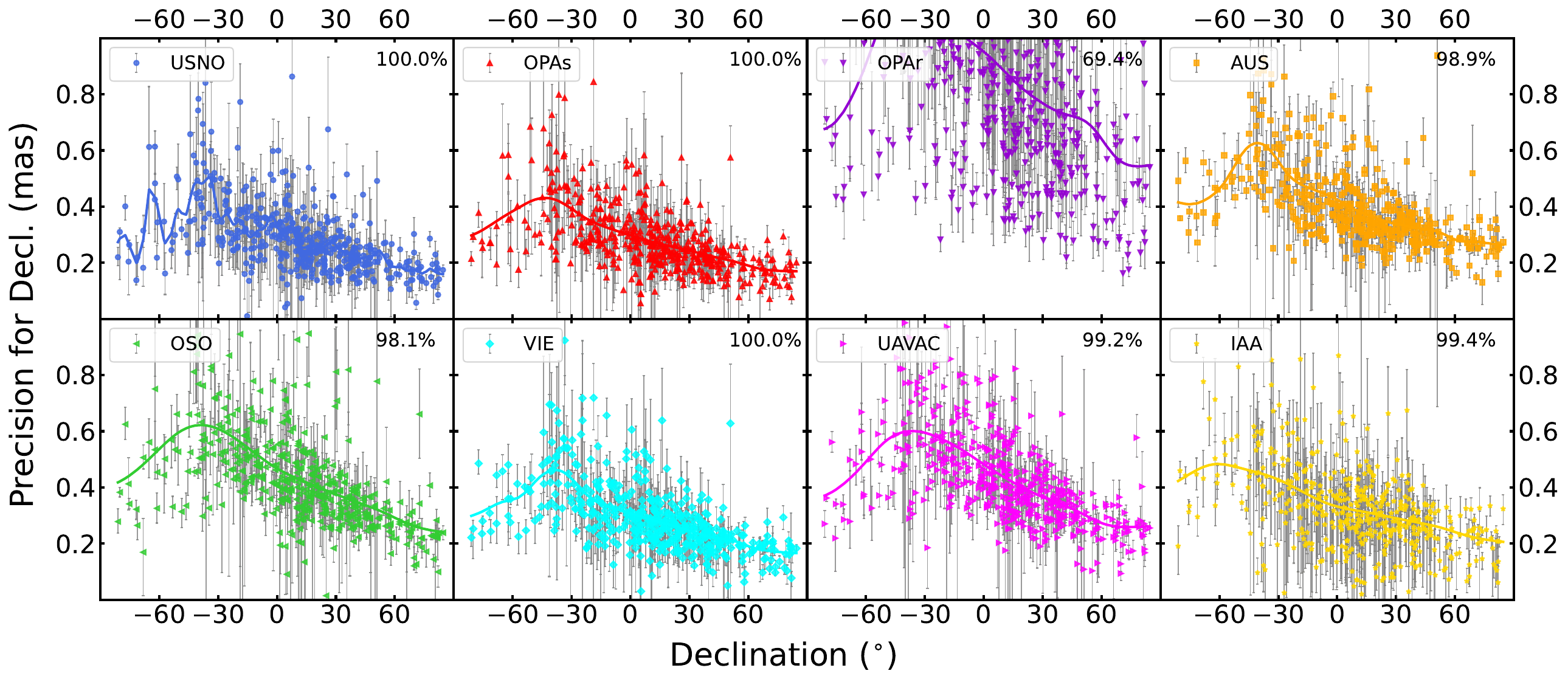}
  \caption{NCH-derived precision for each source from different time-series solutions (color-coded by solution) vs. the declination, shown separately for 481 sources in right ascension (top) and for 474 sources in declination (bottom).
  The error bars denote the 16th and 84th percentiles.
  The number in the top right corner of each panel represents the percentage of data points, which are limited to the range between 0 and 1\,mas.
  The curves show smoothed trends.}
  \label{figA:accuracy scatter declination}
\end{figure}

Figure~\ref{figA:accuracy scatter sessions} shows the NCH-derived positional precision of individual sources as a function of the number of observational sessions in either right ascension or declination.
For most solutions, the precision in both right ascension and declination shows no clear systematic dependence on the number of sessions.
Some exceptions are visible, including a nonlinear trend for OPAr in right ascension and for OPAr, OSO, UAVAC, and IAA in declination, as well as a mild decreasing trend for AUS in right ascension.
The scatter of the precision estimates is larger for sources with fewer sessions, whereas sources with more sessions show more compact distributions.
Consistently, the $68\%$ confidence intervals derived from the bootstrap realizations become narrower as the number of sessions increases.

As shown in Fig.~\ref{figA:accuracy scatter right ascension}, a nonlinear trend of precision on right ascension is observed for either right ascension or declination.
The data points are distributed relatively uniformly over the full range of right ascension.
The smoothed trend curves show weak minor peaks near 4\,h and 14\,h, and shallow troughs around 9\,h and 19\,h.
The relatively sparse sampling of sources is displayed around 21\,h in right ascension.
The OPAr solution shows systematically larger precision values and a wider scatter.

The dependence of the NCH-derived positional precision of source coordinate time series on declination is presented in Fig.~\ref{figA:accuracy scatter declination}.
For most solutions, the precision estimates show a clear declination-dependent pattern.
Larger precision estimates are found around $\delta \simeq -40^{\circ}$ in the Southern Hemisphere, and the estimates generally decrease toward northern declinations.
This peak-like feature and the northward decreasing trend are more pronounced in declination than in right ascension.
In right ascension, the AUS solution shows a more prominent peak around $\delta \simeq -40^{\circ}$ and a stronger nonlinear trend than the other solutions.
By contrast, most other solutions exhibit a simpler, approximately decreasing trend with increasing declination.
The OPAr solution shows systematically larger precision estimates and a broader scatter than the other solutions in both coordinates.
For the smoothed trends shown in Fig.~\ref{figA:accuracy scatter declination}, some fluctuations are visible for the UAVAC solution in right ascension and for the USNO solution in declination, unlike the generally smoother trends of most other solutions.
These fluctuations are likely caused by the large scatter of source-wise precision estimates within the corresponding declination ranges.

\clearpage

\end{appendix}
\end{CJK*}
\end{document}